\documentclass[11pt]{article}
\usepackage[T1]{fontenc}

\usepackage{hyperref}
\usepackage[margin=1in]{geometry}

\usepackage{times}
\usepackage{color}
\hypersetup{
     colorlinks=true,
     linkcolor=blue!50!black,
     citecolor = green!30!black,
     urlcolor=black
}

\usepackage{verbatim}
\usepackage[english]{babel}
\usepackage[utf8]{inputenc}

\usepackage{enumerate}

\usepackage{algorithm}
\usepackage[noend]{algpseudocode}
\usepackage{url}

\usepackage{ifthen}

\usepackage{graphicx}
\usepackage{caption}
\usepackage{subcaption}
\usepackage{tikz}
\usetikzlibrary{calc}
\usetikzlibrary{shapes.geometric}
\usetikzlibrary{arrows}
\usetikzlibrary{decorations.pathreplacing, decorations.pathmorphing}
\usetikzlibrary{patterns}
\usetikzlibrary{backgrounds}
\usetikzlibrary{scopes}
\usetikzlibrary{arrows, automata, positioning}
\usetikzlibrary{decorations.markings}

\usepackage{amssymb,amsmath,amsthm}
\usepackage{mathtools}
\usepackage{thm-restate}

\newtheorem{theorem}{Theorem}[section]
\newtheorem{lemma}[theorem]{Lemma}
\newtheorem{definition}[theorem]{Definition}

\usepackage[colorinlistoftodos,textsize=tiny,textwidth=2cm,color=green!50!gray]{todonotes} 

\usepackage{comment}

\usepackage{xfrac}
\newcommand{\eps}{\varepsilon}
\renewcommand{\epsilon}{\varepsilon}

\newcommand{\dE}{\overrightarrow{E}}

\newcommand{\gapval}{
\frac{16}{9}
}

\newcommand{\mP}{\mathcal{P}}

\ifdefined\DEBUG

\newcommand{\fabr}[1]{\todo[color=cyan!100!black!50]{F: #1}}
\newcommand{\verr}[1]{\todo[color=red!20]{V: #1}}

\newcommand{\jarr}[1]{\todo[color=green!80!black]{J: #1}}
\newcommand{\vera}[1]{\todo[inline,caption=mp,color=red!20]{%
\begin{minipage}{\textwidth}\raggedright \small  #1\end{minipage}%
}}
\else

\newcommand{\fabr}[1]{}
\newcommand{\verr}[1]{}
\newcommand{\jarr}[1]{}

\newcommand{\vera}[1]{}
\fi

\def\Pscr{\mathcal{P}}

\begin{document}

\title{On the Bidirected Cut Relaxation\\ for Steiner Forest}

\author{
Jaros\l aw Byrka\thanks{University of Wroc\l aw,  \href{mailto:jaroslaw.byrka@cs.uni.wroc.pl}%
{jaroslaw.byrka@cs.uni.wroc.pl}. Supported by NCN grant number 2020/39/B/ST6/01641.
}  
\and
Fabrizio Grandoni\thanks{IDSIA, USI-SUPSI,
 \href{mailto:fabrizio.grandoni@idsia.ch}%
{fabrizio.grandoni@idsia.ch}. Partially supported by the SNF Grant $200021\_200731/1$.
}
\and
Vera Traub\thanks{
Research Institute for Discrete Mathematics and Hausdorff Center for Mathematics, University of Bonn,
 \href{mailto:traub@dm.uni-bonn.de}{traub@dm.uni-bonn.de}. Supported by the Deutsche Forschungsgemeinschaft (DFG, German Research
Foundation) – 537750605.
}
}

\date{}

\maketitle

\begin{abstract}
The Steiner Forest problem is an important generalization of the Steiner Tree problem.
We are given an undirected graph with nonnegative edge costs and a collection of pairs of vertices. 
The task is to compute a cheapest forest with the property that the elements of each pair belong to the same connected component of the forest. 
The current best approximation factor for Steiner Forest is $2$, which is achieved by the classical primal-dual algorithm; improving on this factor is a big open problem in the area.  

Motivated by this open problem, we study an LP relaxation for Steiner Forest that generalizes the well-studied Bidirected Cut Relaxation for Steiner Tree. 
We prove that this relaxation has several promising properties. 
Among them, it is possible to round any half-integral LP solution to a Steiner Forest instance while increasing the cost by at most a factor $\frac{16}{9}$.
To prove this result we introduce a novel recursive densest-subgraph contraction algorithm.
\end{abstract}

\section{Introduction}

In the \emph{Steiner Forest} problem we are given an undirected graph $G=(V,E)$ with nonnegative edge costs $c: E \rightarrow \mathbb{R}_{\ge 0}$.
Furthermore, we are given a collection $\mP$ of pairs of vertices. 
The task is to compute a subset of edges $F\subseteq E$ of minimum total cost $c(F):=\sum_{e\in F}c(e)$ such that, for each pair $P=\{s,t\}\in \mP$, the graph $(V,F)$ contains an $s$-$t$ path. 
There is always an optimal solution $F$ where $(V,F)$ is a forest.
We will refer to vertices $R\subseteq V$ that are present in at least one pair from $\mP$ as \emph{terminals}, and call the remaining vertices \emph{Steiner vertices}. 

Steiner Forest generalizes the famous Steiner tree problem, where all terminals need to be connected to each other.
Because Steiner Tree is NP-hard, indeed APX-hard \cite{CC08}, the same applies to Steiner Forest.
The current-best approximation factor for Steiner Forest is $2$, achieved in a classical paper by Agrawal, Klein, and Ravi \cite{AKR91} using the primal-dual method (see also \cite{GW95}). 
Incidentally, this result is also one of the earliest and most famous examples of the primal-dual technique. 
Improving on the $2$-approximation for Steiner Forest is probably among the most prominent open problems in the area of approximation algorithms (see e.g. \cite{williamson2011design}).

The main motivation for our work is a very recent result \cite{BGT24} about the integrality gap  of the Bidirected Cut Relaxation for Steiner Tree (following a long line of research \cite{CDV11,CKP10,hyattdenesik_et_al:LIPIcs.ICALP.2023.79,FKOS16}). 
To define this relaxation, we first \emph{bidirect} all edges, i.e., we replace each undirected edge $\{u,v\}$ with two directed edges $(u,v)$ and $(v,u)$, inheriting the same cost as $\{u,v\}$. 
We let $\dE$ be the resulting set of directed edges.
We also select an arbitrary terminal $r\in R$ as a root. 

The idea behind the relaxation is to consider the equivalent problem of finding a directed Steiner Tree solution $F\subseteq \dE$ where $F$ is the edge set of a tree oriented towards the root $r$.
Any such directed solution $F\subseteq \dE$ has to contain a directed path from each terminal to $r$. 
Equivalently, for every  vertex set $U$ containing some terminal but not $r$, the solution $F$ must contain at least one outgoing edge of $U$.
We write $\delta^+(U)$ to denote the set of outgoing edges of $U$. 
This leads to the Bidirected Cut Relaxation for Steiner Tree, which is defined as follows:
\begin{align}
\min & \sum_{e\in \dE}c(e)\cdot x_e & \tag{Tree-BCR} \label{eq:bcr}\\
\text{s.t.} & \sum_{e\in \delta^+(U)}x_e\geq 1 & \forall U\subseteq V\setminus \{r\}, U\cap R\neq \emptyset. \nonumber
\end{align}
Given a solution $x$ to \ref{eq:bcr}, we denote its cost by $c(x):=\sum_{e\in \dE}c(e)\cdot x_e$. 
\ref{eq:bcr} is one of the oldest and best-studied relaxations for Steiner Tree (see e.g. \cite{GM93}). 
A classical result by Edmonds \cite{E67} shows that it is integral in the special case when $R=V$. 
In this case, Steiner Tree is equivalent to the Minimum Spanning Tree problem.
The best-known lower bound on the integrality gap of \ref{eq:bcr} is only $\frac{6}{5}$  (see \cite{Vicari20},  improving on \cite{BGRS13}). 
This gives hope that \ref{eq:bcr} might have a quite small integrality gap. 
Until very recently however, the best-known upper bound was only $2$. 
In \cite{BGT24} this was improved for the first time, by showing an upper bound of $1.9988$ on the integrality gap of \ref{eq:bcr}.

The latter result motivated us to study the Bidirected Cut Relaxation for Steiner Forest \eqref{eq:forest-bcr}. 
To obtain this relaxation, we bidirect the edges as before. 
Consider any Steiner Forest solution $F$ where each connected component of $(V,F)$ is a tree.
Then we can orient the edges of each such component towards an arbitrary root vertex.
We consider the linear programming relaxation with variables $z^r_P$ for $r\in V$ and $P\in \mathcal{P}$ specifying whether the vertices of the pair $P$ belong to a connected component with root $r$.
Moreover, we have variables $x^r_e$ specifying whether a directed edge $e \in \dE$ is used as part of a connected component with root $r$.
This leads to the following relaxation:

\begin{align}
        \min &\ \sum_{r \in V} \sum_{e \in \overrightarrow{E}} c(e)\cdot x^r_{e} & \tag{Forest-BCR}\label{eq:forest-bcr} \\
        \text{s.t.} \ \ &\ 
         \sum_{r\in V} z^r_{P}\ =\ 1 &\text{ for all } P \in \mathcal{P}  \nonumber \\
        & \sum_{e \in \delta^+(U)} x^r_{e} \ \geq\ z^r_{P} &\text{ for all } r \in V , P\in \mathcal{P},\text{ and }
        U \subseteq V \setminus \{ r \} \text{ with } P \cap U \neq \emptyset  \nonumber  \\
        & x^r_{e} \ \geq\ 0 &\text{ for all }r \in V \text{ and }
        e \in \overrightarrow{E}  \nonumber \\
        & z^r_{P} \ \geq\ 0 &\text{ for all }r \in V \text{ and } P\in \mathcal{P}.  \nonumber
\end{align}

Given a solution $(x,z)$ to \ref{eq:forest-bcr}, we use $c(x):=\sum_{r \in V} \sum_{e \in \overrightarrow{E}} c(e)\cdot x^r_{e}$ as a shortcut for its cost.

The integer program corresponding to \ref{eq:forest-bcr} turned out to perform well in experimental studies \cite{SZM21}\footnote{\ref{eq:forest-bcr} is equivalent to the LP underlying the best performing IP in~\cite{SZM21}.},  but to the best of our knowledge, from a theoretical viewpoint \ref{eq:forest-bcr} has not been studied before.
In this work, we investigate the properties of this LP relaxation, which in the long run might help to improve on the longstanding $2$-approximation algorithm for Steiner Forest.

\subsection{Our Contributions}

One of our main results is an algorithm to round a half-integral solution\footnote{I.e., a solution where the value of each variable is an integer multiple of $\frac{1}{2}$.} to \ref{eq:forest-bcr} while loosing a factor smaller than $2$ in the cost.
Half-integral solutions often play an important role in developing rounding algorithms for network design problems, see e.g.  \cite{carr1998new,boyd202243,10.1145/3357713.3384273}. 
While it remains open whether the integrality gap of \ref{eq:forest-bcr} is less than $2$, the result from next theorem (proof in Section~\ref{sec:half-integral}) gives some supporting evidence that this might be the case. 

\begin{restatable}{theorem}{halfintegral}\label{thm:half-integral}
Given any half-integral solution $(x,z)$ to \ref{eq:forest-bcr}, we can compute in polynomial time  a feasible solution to the associated Steiner Forest instance of cost at most $\gapval \cdot c(x)$.
\end{restatable}

To prove the above result, we provide a simple rounding algorithm which is applicable beyond half-integral solutions.
The half-integrality of $(x,z)$ will be used only in the analysis of the algorithm.

The basic idea of our algorithm is a recursive contraction of densest subgraphs with respect to $x$.
First, we turn the given LP solution $(x,z)$ into a more structured one, applying splitting-off operations as long as this maintains feasibility of the solution.
(Splitting-off means that we reduce the value of variables $x^r_{(u,v)}$ and $x^r_{(v,w)}$ while increasing the value of $x^r_{(u,w)}$ by the same amount.)
Next, we identify a densest subgraph with respect to $x$. 
More precisely, we compute a subset of vertices $W$ that maximizes the sum of the values $x^{r}_{(u,v)}$ with $u,v\in W$, divided by $|W|-1$\footnote{Notice that we do not divide by $|W|$ as it is done in other definitions of density. This difference is crucial for us.}. 
Then we add a minimum spanning tree ${\rm MST}(W)$ on vertex set $W$ to the solution under construction, contract $W$ (updating ${\cal P}$ and $(x,z)$ in the natural way), and continue recursively until each pair is connected. 

We prove that our algorithm yields a solution of cost less than $2 \cdot c(x)$ if the maximum density of a set $W$ with respect to $x$ is always larger than $\frac{1}{2}$.
Moreover, we prove that in the half-integral case, we can achieve a density of at least $\frac{9}{16}$, leading to Theorem~\ref{thm:half-integral}.
\bigskip

Next, we study further basic properties of  \ref{eq:forest-bcr}.
Recall that the pairs $\mathcal{P}$ do not need to be disjoint and a vertex can appear in multiple pairs.
Often the same Steiner Forest instance can be represented in multiple equivalent ways. 
For example, the pairs $\{\{a,b\},\{b,c\}\}$ can be replaced by the pairs $\{\{a,b\},\{a,c\}\}$ without changing the set of feasible solutions. 
We investigate the impact of such representation changes on \ref{eq:forest-bcr}. 

Define the \emph{demand graph} of an instance to be the graph with vertex set $V$ that contains an edge between two vertices $s$ and $t$ if and only if the pair $\{s,t\}$ belongs to $\mathcal{P}$.
All vertices that belong to the same connected component of the demand graph must be connected in any feasible Steiner Forest solution.
Thus if for two instances $((V,E), c, \mathcal{P}_1)$ and $((V,E), c, \mathcal{P}_2)$ the vertex sets of the connected components of the demand graphs are the same, then the two instances are equivalent.
In this case, we also say that $\mathcal{P}_1$ is a different \emph{representation} of $\mathcal{P}_2$ and vice versa.

Even though the feasible solutions of the two instances are identical (and the cost function is the same), it turns out that the value of \ref{eq:forest-bcr} can depend on the representation.

\begin{restatable}{theorem}{differentRepresentations}\label{thm:different_representaions}
There exists an instance $((V,E), c, \mathcal{P}_1)$ of the Steiner Forest problem and a different representation $\mathcal{P}_2$ of $\mathcal{P}_1$ such that the value of \ref{eq:forest-bcr}  is not the same for the two instances $((V,E), c, \mathcal{P}_1)$ and $((V,E), c, \mathcal{P}_2)$.
\end{restatable}

We remark that, on the instances we consider in the proof of Theorem~\ref{thm:different_representaions} (see Section~\ref{sec:different_representations}),  \ref{eq:forest-bcr} has optimal half-integral solutions. 
Hence already in this case, where we show that a good rounding algorithm exists, the representation matters.

Given Theorem~\ref{thm:different_representaions}, one might ask what the best way is to represent a given instance. 
For a given fixed instance, one could simply add all pairs $\{s,t\}$ for which $s$ and $t$ are in the same connected component of the demand graph.
Alternatively, one could also generalize \ref{eq:forest-bcr} by allowing $\Pscr$ to contain arbitrary subsets of $V$ rather than only subsets of size two.
Then one can simply include the vertex set of every connected component of the demand graph in~$\Pscr$.
The latter construction yields a relaxation that is at least as strong as any representation using only pairs.\footnote{
Suppose $\mathcal{P}$ contains a subset $A$, where possibly $|A| >2$, then we can turn any LP solution $(x,z)$ into a feasible LP solution for the instance containing all pairs $P \subseteq A$ by setting $z^r_P \coloneqq z^r_A$ for every such pair and each root $r\in V$.
Hence, the LP relaxation where $A$ is part of $\mathcal{P}$ is at least as strong as any equivalent representation containing pairs $P\subseteq A$ instead of $A$.
}

However, if we want to prove an upper bound on the integrality gap for general instances of Steiner Forest, we need to prove an upper bound for any representation of the instance.
To see this, note that whenever a vertex appears in multiple pairs, we can replace it by several collocated copies. 
This changes neither the value of \ref{eq:forest-bcr} nor the cost of an optimum Steiner Forest solution. 
Of course, one could contract vertices at distance zero upfront, but a similar situation arises if the copies have a very small, but positive distance from each other.
In this new instance, there is only one possible representation of the pairs but the integrality gap of the LP is essentially the same as for the original instance.
\bigskip

We say that a connected component of the demand graph is \emph{trivial} if it contains only a single vertex and we call it \emph{nontrivial} otherwise.
Observe that Steiner Tree instances are instances of Steiner Forest where the demand graph has only one nontrivial connected component. 
In Section~\ref{sec:steiner-tree-case}, we prove that for Steiner Tree instances, the value of \ref{eq:forest-bcr} does not depend on the representation of the pairs $\mathcal{P}$. 
Even more, this value is always equal to the value of \ref{eq:bcr} for the corresponding Steiner Tree instance. 
In particular, the result from \cite{BGT24} implies that such instances have integrality gap strictly smaller than~$2$, namely at most $1.9988$.
\begin{restatable}{theorem}{steiner}\label{thm:steiner}
Let $(G, c, \mathcal{P})$ be an instance of Steiner Forest where the demand graph has only one nontrivial connected component (with vertex set $R$). Then the value of \ref{eq:forest-bcr} is equal to the value of \ref{eq:bcr} for the Steiner Tree instance $(G,c,R)$.
\end{restatable}

Finally, we provide a family of instances showing that the integrality gap of \ref{eq:forest-bcr} is at least $\frac{3}{2}$ (see Section \ref{sec:lowerBound}).

\begin{restatable}{theorem}{integralityGap}\label{thm:lower_bound}
The integrality gap of \ref{eq:forest-bcr} is at least $\frac{3}{2}$.
\end{restatable}

Because \ref{eq:forest-bcr} is a strengthening of the classical undirected LP relaxation of Steiner Forest,\footnote{
To see this, observe that for every feasible solution $(x,z)$ to \ref{eq:forest-bcr}, the vector $\tilde{x}\in \mathbb{R}^E$ defined by $\tilde{x}_{\{u,v\}} \coloneqq \sum_{r\in V} (x^r_{(u,v)} + x^r_{(v,u)})$ is a feasible solution to the undirected LP relaxation.
} 
which has an integrality gap of $2$ \cite{AKR91,GW95,J01}, the integrality gap of \ref{eq:forest-bcr} is at most $2$.
Whether the integrality gap of \ref{eq:forest-bcr} is strictly less than $2$, remains an interesting question for future work.

We remark that for the instances we use to prove Theorem~\ref{thm:lower_bound}, the optimal solutions to \ref{eq:forest-bcr} are not half-integral.
Nevertheless, the rounding algorithm we use to prove Theorem~\ref{thm:half-integral} yields optimal integral solutions for these instances.

\subsection{Related Work}

The \emph{Steiner Network} problem, also known as \emph{Survivable Network Design}, is a generalization of Steiner Forest where we are given pairwise vertex connectivity requirements $\lambda_{u,v}\geq 0$, and the task is to compute a cheapest subgraph of $G$ such that each such pair of vertices $u,v$ is $\lambda_{u,v}$-edge connected. 
In a celebrated result, Jain \cite{J01} obtained a $2$-approximation for this problem using the iterative rounding technique. 
Since then, improving on the $2$-approximation barrier, even just for special cases of Steiner Network, became an important open problem. 
This was recently achieved for some problems in this family, such as Connectivity Augmentation \cite{BGJ23sicomp,CTZ21,TZ21,TZ22,TZ23} and Forest Augmentation \cite{GJT22stoc}. 
Among the special cases for which $2$ is still the best-known factor, we already mentioned the Steiner Forest problem. 
Another interesting special case is the Minimum-Weight 2-Edge Connected Spanning Subgraph problem.

For the Steiner Tree problem several better than $2$ approximation algorithms  are known. 
A series of works based on the relative-greedy approach \cite{Z93,zelikovsky_1996_better,KZ97,PS00} culminated in a $1.55$ approximation \cite{RZ05}. 
The current-best approximation algorithm achieves an approximation guarantee of  $(\ln 4 + \epsilon) \approx 1.39$ \cite{BGRS13} and is based on an iterative randomized rounding approach, applied to the so-called Hypergraphic Cut Relaxation for Steiner tree.
This LP relaxation has  integrality gap at most $\ln 4$ \cite{GORZ12}. 
Recently, the same approximation factor and a new proof of this integrality  gap upper bound has been obtained using local search \cite{TZ22}. 

For Steiner Forest, better-than-2 approximation algorithms are only known in special cases.
In particular, approximation schemes are known for planar and bounded tree-width graphs~\cite{10.1145/2027216.2027219}, and for Euclidean plane instances~\cite{borradaile2015polynomial}. 
A better-than-2 approximation is also known for very dense unit weight graphs~\cite{karpinski2020densesteinerproblemsapproximation}.

For the prize-collecting generalizations of Steiner Tree and Steiner Forest, where some terminals or terminal pairs might be left disconnected by paying a given associated cost, the currently best-known approximation factors are $1.79$ for prize-collecting Steiner Tree \cite{AGHJM24} and $2$ for prize-collecting Steiner Forest~\cite{AGHJM24soda}.

For the directed analogue of the Steiner Forest problem, approximation factors of $O(|R|^{1/2 + \epsilon})$ \cite{chekuri2011set} and $O(|V|^{2/3+\epsilon})$ \cite{berman2013approximation} are known for every fixed $\epsilon > 0$.
For the special case of planar graphs, polylogarithmic approximation factors have been obtained in~\cite{chekuri2024polylogarithmic}.
For the directed analogue of Steiner Tree, the current best approximation guarantees are $|R|^{\eps}$ in polynomial time \cite{CCCDGGL99} and $O(\log^2 |R|/\log\log |R|)$ in quasi-polynomial time \cite{GLL19}.

\subsection{Notation}

For a vertex $v$, we denote by $\delta^-(v)$ and $\delta^+(v)$ the sets of incoming and outgoing edges of $v$, respectively.
Similarly, for a vertex set $U$, we denote by $\delta^-(U)$ and $\delta^+(U)$ the incoming and outgoing edges of $U$, respectively.

For a solution $(x,z)$ to \ref{eq:forest-bcr} and a vertex $r\in V$, we write $c(x^r) \coloneqq \sum_{e\in \overrightarrow{E}} c(e) \cdot x^r_e$.
Moreover, for an edge set $F\subseteq E$, we write $\overrightarrow{F}$ to denote the directed edge set obtained by bidirecting $F$ and define $x^r(F) \coloneqq \sum_{e\in \overrightarrow{F}} x^r_e$ and $x(F) \coloneqq \sum_{r\in V} x^r(F)$.

Finally, for a graph $G=(V,E)$ and a set $W\subseteq V$, we write $G[W]$ to denote the subgraph of $G$ induced by $W$, and we write $E[W]$ to denote the edge set of this subgraph, i.e., the set of all edges from $E$ with both endpoints in $W$.

\section{Rounding half-integral LP solutions}\label{sec:half-integral}

In this section we prove Theorem~\ref{thm:half-integral}, which we restate here for convenience.

\halfintegral*

To prove Theorem~\ref{thm:half-integral} we describe an algorithm that could be applied to round arbitrary solutions $(x,z)$ to \ref{eq:forest-bcr} and then prove that it yields a Steiner Forest solution of cost at most $\gapval c(x)$ if the LP solution $(x,z)$ is half-integral.

\subsection{Description of the algorithm}

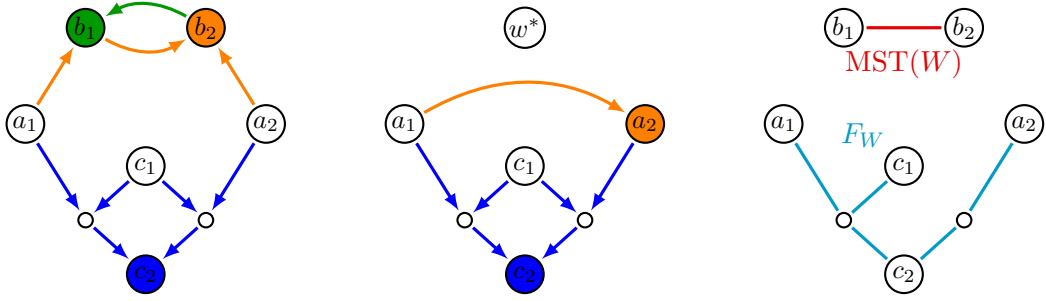
\begin{figure}
\begin{center}
\begin{tikzpicture}[yscale=0.8, xscale=0.8]

\tikzset{terminal/.style={
draw=black,thick, circle,minimum size=1.3em, inner sep=0.5pt, outer sep=1.5pt}
}

\tikzset{steiner/.style={
draw=black,thick, circle,minimum size=0.5em, inner sep=0.5pt, outer sep=1.5pt}
}

\begin{scope}[rotate=90,  xscale=-1]

\begin{scope}[every node/.style={terminal}]
\node (a1) at (0,4) {\small $a_1$};
\node (a2) at (0,0) {\small $a_2$};
\node[fill=green!60!black, fill opacity=0.3, text=black, text opacity=1] (b1) at (-1.6,3) {\small $b_1$};
\node[fill=orange, fill opacity=0.3, text=black, text opacity=1] (b2) at (-1.6,1) {\small $b_2$};
\node (c1) at (0.7, 2) {\small $c_1$};
\node[fill=blue, fill opacity=0.3, text=black, text opacity=1] (c2) at (2.5, 2) {\small $c_2$};
\end{scope}

\begin{scope}[every node/.style={steiner}]
\node (s1) at (1.6, 3) {};
\node (s2) at (1.6, 1) {};
\end{scope}

\begin{scope}[very thick, ->, >=latex]
\begin{scope}[blue]
\draw (a1) -- (s1);
\draw (c1) -- (s1);
\draw (s1) -- (c2);
\draw (c1) -- (s2);
\draw (s2) -- (c2);
\draw (a2) -- (s2);
\end{scope}
\begin{scope}[green!60!black]
\draw[bend left] (b2) to (b1);
\end{scope}
\begin{scope}[orange]
\draw (a1) to (b1);
\draw[bend left] (b1) to (b2);
\draw (a2) to (b2);
\end{scope}
\end{scope}

\end{scope}

\begin{scope}[rotate=90,  xscale=-1, shift={(0, -6.3)}]

\begin{scope}[every node/.style={terminal}]
\node (a1) at (0,4) {\small $a_1$};
\node[fill=orange, fill opacity=0.3, text=black, text opacity=1]  (a2) at (0,0) {\small $a_2$};
\node (b) at (-1.6,2) {\small $w^*$};
\node (c1) at (0.7, 2) {\small $c_1$};
\node[fill=blue, fill opacity=0.3, text=black, text opacity=1] (c2) at (2.5, 2) {\small $c_2$};
\end{scope}

\begin{scope}[every node/.style={steiner}]
\node (s1) at (1.6, 3) {};
\node (s2) at (1.6, 1) {};
\end{scope}

\begin{scope}[very thick, ->, >=latex]
\begin{scope}[blue]
\draw (a1) -- (s1);
\draw (c1) -- (s1);
\draw (s1) -- (c2);
\draw (c1) -- (s2);
\draw (s2) -- (c2);
\draw (a2) -- (s2);
\end{scope}
\begin{scope}[orange]
\draw[bend right] (a1) to (a2);
\end{scope}
\end{scope}

\end{scope}

\begin{scope}[rotate=90,  xscale=-1, shift={(0, -12.6)}]

\begin{scope}[every node/.style={terminal}]
\node (a1) at (0,4) {\small $a_1$};
\node (a2) at (0,0) {\small $a_2$};
\node (b1) at (-1.6,3) {\small $b_1$};
\node (b2) at (-1.6,1) {\small $b_2$};
\node (c1) at (0.7, 2) {\small $c_1$};
\node (c2) at (2.5, 2) {\small $c_2$};
\end{scope}

\begin{scope}[every node/.style={steiner}]
\node (s1) at (1.6, 3) {};
\node (s2) at (1.6, 1) {};
\end{scope}

\begin{scope}[very thick]
\begin{scope}[red!90!black]
\draw (b1) --node[below=4pt] {${\rm MST}(W)$} (b2);
\end{scope}
\begin{scope}[cyan!80!black]
\draw (a1) -- (s1) -- (c1);
\draw (s1) -- (c2) -- (s2) -- (a2);
\node[above left=4pt] at (c1) {$F_W$};
\end{scope}
\end{scope}

\end{scope}

\end{tikzpicture}
\end{center}
\caption{\label{fig:algorithm_example}
The left part of  figure illustrates an LP solution for an instance with pairs 
$\{a_1,a_2\}$, $\{b_1,b_2\}$, and $\{c_1,c_2\}$. 
Every edge $e$ with $x^r_e =\frac{1}{2}$ is drawn in the same color as the vertex $r$.
In the middle, we see the LP solution in the recursive call of Algorithm~\ref{algorithm} (applied to the instance arising from the contraction of $W=\{b_1,b_2\}$) after rerouting and splitting-off.
The solution returned by the algorithm is shown on the right.
}
\end{figure}

We now describe our recursive algorithm, which is summarized in Algorithm~\ref{algorithm}.
See Figure~\ref{fig:algorithm_example} for an example.
If ${\cal P}=\emptyset$, we return the optimum solution, which is $F=\emptyset$.

Otherwise, we compute the metric closure $(\overline{G}, \overline{c})$ of the weighted graph $(G,c)$.
Next, we modify the given LP solution $(x,z)$ into a more structured LP solution for the instance $(\overline{G}, \overline{c}, \mathcal{P})$.
First, we ensure that whenever $z^r_P > 0$ for some vertex $r\in V$ and some pair $P\in \mathcal{P}$, then $r$ is a terminal.
This modification can be done without increasing the cost of the LP solution by a simple rerouting step (see Lemma~\ref{lem:no_steiner_sinks}). 
We remark that we could have required this property in \ref{eq:forest-bcr}, in which case it would be automatically satisfied at the beginning of the algorithm.
However, the property will not be automatically satisfied for the instances on which we call the algorithm recursively.

Then we do splitting-off steps, as long as possible:
If for a vertex $v\in V$, a root $r\in V$, and edges $(u,v)$ and $(v,w)$ we can reduce $x^r_{(u,v)}$ and $x^r_{(v,w)}$ by $\epsilon > 0$ and increase $x^r_{(u,w)}$ by $\epsilon$ while maintaining the feasibility of our LP solution, then we do so for the largest possible such $\epsilon$.
Because the cost function $\overline{c}$ is metric, this does not increase the cost of our LP solution.
We then obtain an LP solution $(x,z)$ with the following two properties:
\begin{enumerate}[(a)]\itemsep1pt
\item We have $z^r_P = 0$ for all $r\in V\setminus R$ and all $P\in \Pscr$.
\item No splitting-off step is possible while maintaining the feasibility of $(x,z)$.
\end{enumerate}
We call an LP solution with these two properties \emph{well structured}.

\begin{algorithm}[t]
      \caption{Recursive algorithm to round a solution $(x,z)$ to \ref{eq:forest-bcr} for a given instance $(G,c,{\cal P})$.\label{algorithm}}
    \begin{algorithmic}[1]
    \State If ${\cal P}=\emptyset$, \textbf{return} $\emptyset$.
        \State Let $(\overline{G},\overline{c})$ be the metric closure of $(G,c)$.
        \State Replace $(x,z)$ with a well structured solution for the instance $(\overline{G}, \overline{c},\mathcal{P})$ without increasing the cost.
            \State Compute a densest subgraph of $G$ with respect to edge weights $x$, i.e., compute a set $W\subseteq V$ maximizing ${\rm density}_x(W):=\frac{x(E[W])}{|W|-1}$. 
        \State Compute a minimum spanning tree ${\rm MST}(W)$ of $\overline{G}[W]$ with respect to the costs $\overline{c}$.
        \State Call the algorithm recursively on the instance and LP solution that arise by contraction of $W$ and let $F_W$ be the resulting edge set.
        \State Let $F$ result from $F_W \cup {\rm MST}(W)$ by replacing every edge $\{v,w\}$ by a shortest $v$-$w$ path in $(G,c)$.
        \State \textbf{Return} $F$.
    \end{algorithmic}
\end{algorithm}

Having a well structured LP solution for the instance $(\overline{G}, \overline{c}, \mathcal{P})$, we compute a densest subgraph of $\overline{G}=(V,E)$ with respect to the following notion of density:
\begin{definition}
For $W \subseteq V$ with $|W| \ge 2$, we define the density of $W$ with respect to $x$ as
\[
 {\rm density}_x(W) \coloneqq \frac{x(E[W])}{|W|-1}.
\]
\end{definition}
We compute $W \subseteq V$ with $|W|\ge 2$ maximizing $ {\rm density}_x(W)$.
Such a set $W$ can be found in polynomial time by a reduction to submodular function minimization; see Lemma~\ref{lem_find_dense} in Section~\ref{sec:poly_time}.

Next we compute a minimum cost spanning tree ${\rm MST}(W)$ in the graph $\overline{G}[W]$ induced by $W$ with respect to edge costs $\overline{c}$.
Then we call the algorithm recursively on the instance obtained by the contraction of $W$ into a vertex $w^*$.
We identify the edges in the contracted instance with the corresponding edges in the original instance. 
If parallel edges would arise by the contraction, we keep only one edge with minimal cost.
Edges with both endpoints in $W$ are no longer present after contraction.

When contracting the set $W$, we adapt the set of pairs in the natural way: we remove every pair $P\in \mathcal{P}$ with $P\subseteq W$; in every pair $\{s,t\}$ with $s\in W$ and $t\notin W$, we replace $s$ with $w^*$.
We also transform the LP solution $(x,z)$ in the canonical way as follows.
We set $z^{w^*}_P \coloneqq \sum_{w\in W}  z^{w}_P$ for each $P\in {\cal P}$.
For all vertices $r, v\in V\setminus W$, we set $x^r_{(v,w^*)} \coloneqq \sum_{w \in W} x^r_{(v,w)}$ and similarly $x^r_{(w^*, v)} = \sum_{w \in W} x^r_{(w, v)}$.
Moreover, for each vertex $v\in V\setminus W$, we set $x^{w^*}_{(v,w^*)} \coloneqq \sum_{w \in W} \sum_{u\in W} x^w_{(v,u)}$ and similarly $x^{w^*}_{(w^*,v)} \coloneqq \sum_{w \in W} \sum_{u\in W} x^w_{(u, v)}$.

Then we take the union of ${\rm MST}(W)$ and the edge set $F_W$ obtained from the recursive call to obtain a solution to the instance $(\overline{G}, \overline{c}, \mathcal{P})$.
Finally, we replace every edge $\{v,w\}$ by a shortest $v$-$w$ path in the original graph $G$ with respect to the cost function $c$.
In this way, we obtain a Steiner Forest solution to the original instance $(G,c,\mathcal{P})$.

\subsection{Polynomial runtime of the algorithm}\label{sec:poly_time}

In this section we explain how to perform the different steps of Algorithm~\ref{algorithm} in polynomial time.
We first prove how to compute a vertex set $W$ maximizing ${\rm density}_x(W)$.

\begin{lemma} \label{lem_find_dense}
    Given a rational solution $(x,z)$ to \ref{eq:forest-bcr}, we can in polynomial time find a set $W\subseteq V$ with $|W|\ge 2$ that maximizes ${\rm density}_x(W) = \frac{x(E[W])}{|W|-1}$.
\end{lemma}

\begin{proof}
First, we prove that given a number $\gamma \ge 0$, we can in polynomial time check whether there exists a set $W$ with $|W| \ge 2$ and ${\rm density}_x(W) \ge \gamma$ and if this is the case, we can find such a set $W$.

To do so we will guess a pair of vertices $u$ and $v$ that will be part of $W$, and only check if there is a set $W$ containing both $u$ and $v$ with ${\rm density}_x(W) \ge \gamma$.
Let $\tilde{V} = V \setminus \{u,v\}$ and for $\tilde{U}\subseteq \tilde{V}$, let $U \coloneqq \tilde{U}\cup \{u,v\}$.
Then we consider the set function $h : 2^{\tilde{V}} \rightarrow \mathcal{R}$ defined by
 \[
    h(\tilde{U}) \coloneqq \delta \cdot (|U| -1) - x(E[U]).
\]
Observe that the function $h$ is submodular.  
Therefore, there is a strongly polynomial time algorithm to compute a set $\tilde{U}$ minimizing the function $h$ \cite{schrijver2000combinatorial,iwata2001combinatorial}.

    In case $h(\tilde{U})\leq 0$, we have ${\rm density}_x(U) \ge \gamma$.
Otherwise, there is no set $W$ with ${\rm density}_x(W) \ge \gamma$ and $\{u,v\} \subseteq W$ because otherwise $h(W\setminus\{u,v\}) < 0$.
By performing this computation for each possible choice of the vertices $u$ and $v$, we can indeed find a set $W$ with $|W|\ge 2$ and ${\rm density}(W)\ge \gamma$ or decide that no such set $W$ exists.

We now use binary search to (approximately) guess the optimal density $\gamma^*$, i.e., $\gamma^* \coloneqq \max \{ {\rm density}_x(W) : W\subseteq V\text{ with }|W|\ge 2\}$.
We have $0 \le \gamma^* \le x(E)$.
Using the algorithm described above we can decide for any given value $\gamma $ whether $\gamma^*\leq \gamma$. 
Let $M\in \mathbb{N}$ be smallest possible such that the vector $M \cdot x$ has only integral entries.
Using binary search, we can find values $a,b\in [0,x(E)]$ with $\gamma^*\in [a,b]$ and $b-a \le \frac{1}{M \cdot |V|^2}$.
This requires $\log_2( M \cdot x(E)) + \log_2(|V|)$ iterations,  which is polynomially bounded in the encoding length of the input.
(Note that the encoding length of $x$ is  $\Omega(\log( M \cdot x(E)))$.)

We then use the algorithm described above to compute a set $W\subseteq V$ with $|W| \geq 2$ and ${\rm density}_x(W) \geq a$.
We claim that ${\rm density}_x(W) = \gamma^*$.
To see this, suppose we have ${\rm density}_x(W) < \gamma^*$ and let $W^* \subseteq V$ with $|W^*|\geq 2$ and ${\rm density}_x(W^*) = \gamma^*$.
Then using that $M\cdot x(E[W^*]) \in \mathbb{N}$ and $M\cdot x(E[W]) \in \mathbb{N}$, we get
\begin{align*}
M \cdot (b-a) \ \geq&\ M \cdot \bigl({\rm density}_x(W^*) - {\rm density}_x(W) \bigr) \\[2mm]
=&\ \frac{M \cdot x(E[W^*])}{|W^*| -1} - \frac{M \cdot x(E[W])}{|W| -1}  \\[2mm]
 \geq&\ \frac{1}{(|W^*| -1) \cdot (|W| -1)} \\[2mm]
  >&\ \frac{1}{|V|^2},
\end{align*}
a contradiction.
\end{proof}

Next we prove that we can ensure that whenever $z^r_P > 0$ for some vertex $r\in V$ and some pair $P\in \mathcal{P}$, then $r$ is a terminal.

\begin{lemma}\label{lem:no_steiner_sinks}
Given a feasible solution $(x,z)$ to \ref{eq:forest-bcr}, we can in polynomial time compute a feasible solution $(\tilde{x},\tilde{z})$ of at most the same cost where $\tilde{z}^r_P = 0$ for all $r\in V\setminus R$ and all $P\in \mathcal{P}$.
Furthermore, if $(x,z)$ is half-integral, then $(\tilde{x},\tilde{z})$ is also half-integral.
\end{lemma}

\begin{proof}
Consider a vertex $r \in V \setminus R$ such that $z^r_P > 0$ for some pair $P$. 
Let $P^*$ be a pair maximizing $z^r_{P^*}$ and let $v\in P^*$.
If we interpret each value $x^r_e$ as capacity of the edge $e$, than there exists a flow $f$ from $v$ to $r$ of value $z^r_{P^*}$ satisfying these capacities.
If $(x,z)$ is half-integral, we can choose $f$ to be half-integral.
We will modify $x^r$ by reverting this flow and then changing the root from $r$ into $v$. 
More specifically, we first define a vector $\bar x^r \in \mathbb{R}^{\overrightarrow{E}}$ by
\[
    \bar x^r_{(u,w)} \ =\ x^r_{(u,w)} - f_{(u,w)} + f_{(w,u)}.
\]
Then we define $(\tilde{x},\tilde{z})$ by setting for each edge $e \in \overrightarrow{E}$
\begin{align*}
\tilde{x}^r_e =&\  0  &  \\
\tilde{x}^v_e =&\ x^v_e + \bar x^r_e &\\
\tilde{x}^u_e =&\ x^u_e & & \text{ for all } u\in V \setminus\{v,r\}
\end{align*}
and setting for each pair $P\in \mathcal{P}$
\begin{align*}
\tilde{z}^r_P =&\  0  \\
\tilde{z}^v_P =&\ z^v_P +  z^r_P   \\
\tilde{z}^u_P =&\ z^u_P & &\text{ for all } u\in V \setminus\{v,r\} .
\end{align*}
Then we have $c(\tilde{x}) = c(x)$.

Next, we prove that $(\tilde{x}, \tilde{z})$ is a feasible solution to \ref{eq:forest-bcr}.
Because the flow $f$ satisfies the capacities $x^r$, the vector $\overline{x}^r$ has only nonnegative entries and hence the same holds for $\tilde{x}$.
For every pair $P \in \mathcal{P}$, we have $\sum_{r\in V} \tilde{z}^r_P = \sum_{r\in V} z^r_P = 1$.
Now consider a set $U\subseteq V\setminus \{v\}$ and a pair $P\in \mathcal{P}$.
If $r\notin U$, we have $f(\delta^+(U))- f(\delta^-(U)) = 0$ and therefore
\[
    \overline{x}^r(\delta^+(U)) = x^r(\delta^+(U)) - f(\delta^+(U))+ f(\delta^-(U)) = x^r(\delta^+(U))  \geq z^r_P.
\]
If $r\in U$, we have $f(\delta^+(U))- f(\delta^-(U)) = - z^r_{P^*}$ and thus
\[
    \overline{x}^r(\delta^+(U)) = x^r(\delta^+(U)) - f(\delta^+(U))+ f(\delta^-(U)) \geq  z^r_{P^*} \geq z^r_P.
\]
Thus, in both cases we have $ \overline{x}^r(\delta^+(U)) \geq z^r_P$, which together with $x^v(\delta^+(U)) \geq z^v_P$ implies $\tilde{x}^v(\delta^+(U)) \geq \tilde{z}^v_P$.
All other constraints of \ref{eq:forest-bcr} for $(\tilde{x}, \tilde{z})$ follow directly from the fact that $(x,z)$ satisfies these constraints.
We conclude that $(\tilde{x}, \tilde{z})$ is a feasible solution to \ref{eq:forest-bcr}.

We apply the above transformation, which can be performed in polynomial time, to each vertex $r\in V\setminus R$ for which $z^r_P > 0$ for some pair $P\in \mathcal{P}$.
\end{proof}

Finally, we prove that our algorithm has polynomial runtime and we can maintain half-integrality of the LP solution.

\begin{lemma}\label{lem:runtime_and_halfintegrality}
Algorithm~\ref{algorithm} has polynomial runtime.
If the given LP solution is half-integral, we can ensure that the LP solution remains half-integral throughout the algorithm.
\end{lemma}

\begin{proof}
Because $|W|\ge 2$, in every recursive call of the algorithm the number of vertices decreases by at least one. 
Hence, the total number of recursive calls is at most $|V|$.
Computing the set $W$ can be done in polynomial time by Lemma~\ref{lem_find_dense}.
The only step of the algorithm where it is not obvious that it can be performed in polynomial time is replacing $(x,z)$ by a well structured LP solution without increasing the cost.

In order to do so, we first apply Lemma~\ref{lem:no_steiner_sinks}.
Next, we iterate over all roots $r\in R$ and all vertices $v\in V$.
Then for each $r\in R$ and each $v\in V$, we apply the following splitting-off operation to each pair of edges $(u,v)\in \delta^-(v)$ and $(v,w)\in \delta^+(v)$.
We compute the maximum $\epsilon \geq 0$ such that decreasing $x^r_{(u,v)}$ and $x^r_{(v,w)}$ by $\epsilon$ and increasing $x^r_{(u,w)}$ by $\epsilon$ maintains a feasible solution to \ref{eq:forest-bcr} (see below for how to compute this value $\epsilon$).
Then we decrease $x^r_{(u,v)}$ and $x^r_{(v,w)}$ by $\epsilon$ and increase $x^r_{(u,w)}$ by $\epsilon$.
Observe that by this splitting-off operation the value $x^r(\delta^+(U))$ decreases by $\epsilon$ for each set $U$ with $v\in U$ and $U \subseteq V \setminus \{u,w\}$.
For all other sets $U\subseteq V$, the value of $x^r(\delta^+(U))$ remains unchanged.
Therefore, the maximum feasible value $\epsilon$ is the minimum of $x^r_{(u,v)}$,  $x^r_{(v,w)}$,  and the minimum of $x^r(\delta^+(U)) - z^r_P$ over all pairs $P\in \mathcal{P}$ and all sets $U\subseteq V\setminus \{r,u,v\}$ with $P\cap U \neq \emptyset$ and $v\in U$.
To compute this maximum value $\epsilon$ we can enumerate over all pairs $P\in \mathcal{P}$, guess a vertex in $P\cap U$ and then find a set $U$ attaining the minimum by a minimum-cut computation.
If $(x,z)$ is half-integral, $x^r(\delta^+(U)) - z^r_P$ is half-integral for all vertex sets $U$ and pairs $P$, implying that $\epsilon$ is half-integral.

It remains to prove that for each root $r\in R$ it suffices to apply splitting-off at each vertex $v$ only once.
To see this, recall that for every $U\subseteq V$, the value $x^r(\delta^+(U))$ can never increase during the splitting-off process.
Thus, the only reason for why a splitting-off step with a positive value $\epsilon$ might be possible after doing splitting-off steps at other vertices is that the value on an incoming edge $(u,v)$ of $v$ or an outgoing edge $(v,w)$ of $v$ increased.
The second case is symmetric to the first one; hence we focus on the first.
Consider a sequence of edges $(u_1, u_2), (u_2,u_3), \dots , (u_{l-1}, u_l), (u_l,v)$, and $(v,w)$ where we considered the vertex $v$ before the vertices $u_1, \dots, u_l$ when applying splitting-off operations.
Suppose that after splitting-off operations at $u_2, \dots, u_l$, we can now reduce $x^r_{(u_1,v)}$ and $x^r_{(v,w)}$ by $\epsilon > 0$ when increasing $x^r_{(u_1,w)}$ by $\alpha$.
Then we could have already reduced $x^r_{(u_l,v)}$ and $x^r_{(v,w)}$ and increased $x^r_{(s,w)}$ by $\alpha$ before the splitting-off operations at $u_2,\dots, u_l$ contradicting the fact that we chose the maximum possible value $\epsilon$ when we considered the vertex~$v$ and edges $(u_l,v)$ and $(v,w)$.
\end{proof}

\subsection{Analysis of the approximation ratio}\label{sec:analysis}
 
We next analyze the approximation ratio of the algorithm and prove that we obtain a solution of cost at most $\gapval \cdot c(x)$ if $(x,z)$ is half-integral.

First, we discuss how we can bound the cost ${\rm mst}(W)$ of the 
the minimum spanning tree ${\rm MST}(W)$ in $\overline{G}[W]$.
We denote by $x|_{E[W]}$ the vector $x$ restricted to directed edges with both endpoints in $W$ and therefore write  $c(x|_{E[W]}) = \sum_{v,w\in W} c(e) \cdot \sum_{r\in V} x^r_{(v,w)}$.

\begin{restatable}{lemma}{mstcost}\label{lem:bound_mst_cost}
Let $(x,z)$ be a solution to \ref{eq:forest-bcr} and let $W\subseteq V$ with $|W|\ge 2$ maximizing ${\rm density}_x(W)$.
Then 
\[
{\rm mst}(W) \le \frac{c(x|_{E[W]})}{{\rm density}_x(W)}.
\]
\end{restatable}

\begin{proof}
We consider the partition LP for the minimum spanning tree problem in $\overline{G}[W]$, which can be described as follows.
We write $\Pi$ to denote the family of all partitions of $W$.
For a partition $\pi\in \Pi$ with $|\pi|$ parts, we write $\delta(\pi)$ to denote the set of edges of $\overline{G}[W]$ with endpoints in different parts of the partition $\pi$.
Then the partition LP is the following linear program:
\begin{align}\label{eq:partition-lp}\tag{Partition-LP}
        \min &\; \sum_{e \in E[W]}  c(e) \cdot y_{e} & \\
        \text{s.t.} &\ 
        y(\delta(\pi))\geq |\pi|-1 & & \text{ for all } \pi \in \Pi;\nonumber \\
        &\ y_{e} \geq 0 & & \text{ for all }e\in E[W].\nonumber
\end{align}
By Theorem~50.8 in~\cite{schrijver2003combinatorial},the optimum value of \ref{eq:partition-lp} is equal to ${\rm mst}(W)$.
We claim that the vector $y$ given by
\[
y_{\{v,w\}} \coloneqq \frac{1}{{\rm density}_x(W)} \cdot \sum_{r\in R} \Bigl( x^r_{(v,w)} + x^r_{(w,v)} \Bigr)
\]
for every edge $\{v,w\}\in E[W]$ is a feasible solution to \ref{eq:partition-lp}.
Because $c(y) = \frac{c(x|_{E[W]})}{{\rm density}_x(W)}$, this implies Lemma~\ref{lem:bound_mst_cost}.

To prove that $y$ is indeed a feasible solution to \ref{eq:partition-lp}, we first observe $y(E[W]) = |W| -1$ by the definition of ${\rm density}_x(W)$.
Now consider a partition $\pi$ of $W$.
Because $W$ is maximizing the density with respect to $x$ among all vertex sets with at least two elements, we have
${\rm density}_x(W) \ge \frac{x(E[U])}{|U|-1}$ for each set $U\in \pi$ with $|U|\geq 2$.
Therefore, we have $y(E[U]) = \frac{x(E[U])}{{\rm density}_x(W)} \leq |U| -1$ for each $U\in \pi$ with $|U|\geq 2$.
Because, $y(E[U]) \leq |U| -1$ clearly also holds for each set $U\in \pi$ with $|U|=1$, we get
\[
y(\delta(\pi)) \  =\ y(E[W]) - \sum_{U \in \pi} y(E[U]) \ \geq \  |W| - 1 -  \sum_{U \in \pi} (|U|-1) \ =\ |\pi| - 1.
\]
\end{proof}

The next step our analysis is to prove that for a well structured half-integral LP solution $(x,z)$, a sufficiently dense subgraph with respect to $x$ does always exist:

\begin{restatable}{lemma}{density} \label{lem:dense_half_integral}
    Let $(x,z)$ be a well structured solution to \ref{eq:forest-bcr} for an instance with $|\mathcal{P}| \geq 1$.
    If $(x,z)$ is half-integral, then there exists a set $W\subseteq V$ with $|W| \geq 2$ and ${\rm density}_x(W) \geq \frac{9}{16}$.
\end{restatable}

The proof of Lemma~\ref{lem:dense_half_integral} is the only part of our analysis where we use the half-integrality of the LP solution.

First, we will observe that we may assume that reducing any variable $x^r_e$ with $r\in V$ and $e\in \overrightarrow{E}$ by any $\epsilon > 0$ will not maintain a feasible solution to \ref{eq:forest-bcr}.
We call a well structured LP solution with this property \emph{fully reduced}.
If $(x,z)$ is not fully reduced, we can reduce the value of variables $x^r_e$ to obtain a fully reduced solution $(x',z)$.
Note that if we can reduce a variable $x^r_e$ by some $\epsilon > 0$, then we can reduce it by a positive half-integral value.
Thus, if we consider the variables $x^r_e$ one by one and reduce each of them as much as possible, then half-integrality will be maintained.
Then the fact that  ${\rm density}_{x'}(W) \le {\rm density}_x(W)$ for every set $W\subseteq V$ with $|W|\ge 2$ implies that it suffices to prove the statement for the fully reduced solution $(x',z)$. 
Note that the reduction of a variable $x^r_e$ also maintains the property that LP solution is well structured.

We will now exploit the assumption that we work with half-integral solutions
to give a combinatorial argument for that there always exists a dense subgraph.
We consider the \emph{projection multi-graph} $\hat{G}=(V, \hat{E})$ of half-integral solution $(x,z)$ that contains $2 \cdot \sum_{r\in V} ( x^r_{(v,w)} + x^r_{(w,v)} )$ copies of each edge $\{v,w\}$.

We call a vertex a \emph{support vertex} of $(x,z)$ if it has at least one incident edge in $\hat G$.
We say that a vertex is a \emph{high-degree} vertex if its has at least $3$ incident edges in $\hat G$.
Otherwise, it is a \emph{low-degree} vertex.

\begin{lemma} \label{lem_high_degree}
    Let $(x,z)$ be a fully reduced well structured solution and let $\hat{G}$ be its projection multi-graph.
    Then for every support vertex $v$, at least one of the following holds:
 \begin{enumerate}[(i)]
     \item\label{item:parallel} there is a vertex $w \in V$ such that $\hat{G}$ contains two parallel edges $\{v,w\}$;
     \item\label{item:high-degree} either $v$ itself is a high-degree vertex, or one of its neighbors in $\hat{G}$ is a high-degree vertex.
 \end{enumerate}
 Moreover, every support vertex has degree at east two in $\hat G$.
\end{lemma}

\begin{proof}
We first prove the following claim: \\

\noindent     {\bf Claim 1.}  
Every Steiner vertex that is a support vertex is a high-degree vertex.\\
   
To prove Claim~1, consider a Steiner vertex $v$ that is a support vertex.
Because $(x,z)$ is fully reduced, for every root $r\in V$ we either have $x^r(\delta^-(v)) = 0 = x^r(\delta^+(v))$ or we have $x^r(\delta^-(v))> 0$ and $x^r(\delta^+(v)) > 0$.
The latter case applies for at least one root $r\in V$ because $v$ is a support vertex.
If $x^r(\delta^-(v))=x^r(\delta^+(v))= \frac{1}{2}$, then $v$ has an incoming edge $(u,v)$ with $x^r_{(u,v)}=\frac{1}{2}$ and an outgoing edge $(v,w)$ with $x^r_{(v,w)}=\frac{1}{2}$ and we have $x^r_e = 0$ for all other directed edges incident to $v$.
Then a splitting-off operation increasing $x^r_{(u,w)}$ by $\frac{1}{2}$ and decreasing $x^r_{(u,v)}$ and $x^r_{(v,w)}$ to zero would maintain feasibility of the LP solution, contradicting the fact that $(x,z)$ is well structured.
Therefore, $x^r(\delta^-(v)) + x^r(\delta^+(v)) \ge \frac{3}{2}$, implying that $v$ is a high-degree vertex.

Next, we prove the following sufficient condition for \eqref{item:high-degree}:\\
  
\noindent    {\bf Claim 2.}
  If $v$ is a terminal and $x^r_e > 0$ for some $r\in V\setminus \{v\}$ and $e\in \delta^-(v)$, then $v$ is a high-degree vertex.  \\
    
    To see that Claims 2  holds, consider a terminal $v\in R$ and a pair $P\in \mathcal{P}$ with $v\in P$.
    Then by the feasibility of $(x,z)$ for \ref{eq:forest-bcr}, we have:
    \[
       \sum_{r \in V} x^r(\delta^+(v)) \geq \sum_{r \in V\setminus\{v\}}z^r_P  =1 - z^v_P,
    \]
    and
    \[
      x^v(\delta^-(v))  \geq z^v_P.
    \]
Therefore, if $x^r_e > 0$ for some $r\in V\setminus \{v\}$ and $e\in \delta^-(v)$,  we get $x(\delta(v)) \ge x^r_e  +  x^v(\delta^-(v))  +  \sum_{r \in V} x^r(\delta^+(v)) > 1$, implying that $v$ is a high-degree vertex.

    Having proved the two claims, we now use them to prove the statement of Lemma~\ref{lem_high_degree}.
    If a support vertex is a Steiner vertex, then it has degree at least three in $\hat G$ by Claim~1.
    For a support vertex that is a terminal, consider a pair $P\in \mathcal{P}$ with $v\in P$.
Then we have $x^v(\delta^-(v)) \geq z^v_P$ and for every root $r\in R\setminus\{v\}$, we have $x^r(\delta^+(v)) \geq z^r_P$.
 Therefore, $x(\delta^+(v)) + x(\delta^-(v)) \geq \sum_{r\in R} z^r_P = 1$.
  We conclude that every support vertex has degree at least two in $\hat G$.
    
    Let now $v\in V$ be a support vertex  and suppose $v$ does not satisfy \eqref{item:high-degree}, i.e., $v$ is a low-degree vertex and also all its neighbors in $\hat G$ are low-degree vertices.
    Because $v$ is a low-degree vertex, by Claim 1, it must be a terminal (and also its neighbors in $\hat G$ must be terminals).
    Hence, there exists a pair $P \in \mathcal{P}$ such that $v \in P$. 
    
    We observe that for every root $r\in V$, there is a flow $f \in \mathbb{R}^{\overrightarrow{E}}$ sending $z^r_P$ units of flow from $v$ to $r$ with $f_e \le x^r_e$ for all $e\in \overrightarrow{E}$ (by the constraints of \ref{eq:forest-bcr} and the max-flow min-cut theorem).
    In this case, we will say that $v$ can send $z^r_P$ units of flow  to $r$ in $x^r$.

    We now distinguish two cases.\\
    
     \noindent\textbf{Case 1:} For every pair $P\in \mathcal{P}$ with $v\in P$, there exists a vertex $r \notin P$ such that $z_P^r > 0$. 
       
    We fix a pair $P\in \mathcal{P}$ with $v\in P$ and let  $r \notin P$ such that $z_P^r > 0$. 
    Then we have $x_{(v,u)}^r > 0$ for some vertex $u\in V$.
     If $u \neq r$, then  $u$ is a high-degree vertex by Claim 2. 
     In the remaining part of the argument, we will hence assume $u = r$. 
     In particular, $u$ is a terminal because $(x,z)$ is well structured.
     Because $v$ is not a high-degree vertex, by Claim~2, we have $x^u_e = 0$ for all $e\in \delta^-(v)$.
     Using that the vertex $v' \in P\setminus\{v\}$ needs to route some flow to $r=u$ in $x^u$ (but this flow can never enter $v$), there must be a vertex $w \neq v$ with $x^u_{(w,u)} > 0$.

    Because all neighbors of $v$, and thus in particular $u$, are low-degree vertices, we have $x(\delta(u)) = 1$.
    Using  $x_{(v,u)}^r > 0$ and $x^u_{(w,u)} > 0$ (and $x$ is half-integral), this implies $x(\delta^+(u)) = 0$.
    Because $u$ is a terminal, there is a pair $\tilde{P} \in \mathcal{P}$ containing $u$ and because of $x(\delta^+(u)) = 0$ we must have $z^u_{\tilde{P}} = 1$.
    Let $u'\in \tilde{P}$ with $u' \neq u$.
    Because we are in Case~1, we must have $v \neq u'$.
   The terminal $u' \in \tilde{P}$ must send one unit of flow to $u$ in $x^u$ and because $x^u_e = 0$ for all $e\in \delta^-(v)$, this flow cannot use the edge $(v,u)$.
   But then $x^u(\delta^-(u)) \ge  1+ x_{(v,u)}^u > 1$ and hence $u$ is a high-degree vertex.
   \\

     \noindent\textbf{Case 2:} There exists a pair $P\in \mathcal{P}$ with $v\in P$ such that for all $r \notin P$ we have $z_P^r = 0$. 
      
     Let $P\in \mathcal{P}$ with $v\in P$ such that for all $r \notin P$ we have $z_P^r = 0$ and let $w\in P \setminus \{v\}$.
     Then we have $P= \{v, w\}$ and $z^v_P + z^{w}_P = 1$.
     If $x^v_{(w,v)} \ge z^v_P$ and $x^{w}_{(v,w)} \ge z^{w}_P$, then the projection multi-graph $\hat{G}$ contains two parallel edges between $v$ and $w$.
     Therefore, we have $x^{w}_{(v,w)} < z^{w}_P$ or $x^{v}_{(w,v)} < z^{v}_P$.
     
     First consider the case $x^{w}_{(v,w)} < z^{w}_P$. 
     Then there must be $z^{w}_P$ units of flow from $v$ to $w$ in $x^w$, but not all of the flow can be on the edge $(v,w)$.
     Hence, we must have $x^w_{(v,u)} > 0$ for some edge $(v,u)$ with $u\neq w$.
     Then Claim~2 implies that $u$ is a high-degree vertex.
     
     Now consider the remaining case $x^{v}_{(w,v)} < z^{v}_P$.
     Then there must be $z^{v}_P$ units of flow from $w$ to $v$ in $x^v$, but not all of the flow can be on the edge $(w,v)$.
     Hence, we must have $x^v_{(u,v)} > 0$ and $x^v_e > 0$ for some vertex $u\neq w$ and some edge $e\in \delta^-(u)$.
     Then $u$ is a high-degree vertex by Claim~2.
\end{proof}

We are now ready to prove Lemma~\ref{lem:dense_half_integral}.

\begin{proof}[Proof of Lemma~\ref{lem:dense_half_integral}.]
We assume without loss of generality that $(x,z)$ is fully reduced.
We observe that the definition of the projection multi-graph $\hat G=(V, \hat E)$ implies that for every set $W\subseteq V$ with $|W|\geq 2$, the density with respect to $x$ is
${\rm density}_x(W) =\frac{|\hat E[W]|}{2 \cdot (|W|-1)}$, where $\hat E[W]$ is the set of edges of $\hat G$ with both endpoints in $W$.

We apply Lemma~\ref{lem_high_degree}.
If for some support vertex $v$, property~\eqref{item:parallel} is satisfied, i.e., there exists a vertex $w$ such that $\hat G$ contains at least two parallel edges between $v$ and $w$, then we set $W \coloneqq \{v,w\}$ and get ${\rm density}_x(W) \geq 1$.

Hence, we may now assume that property~\eqref{item:parallel} is not satisfied for any support vertex $v$.
Then by Lemma~\ref{lem_high_degree}, every support vertex satisfies~\eqref{item:high-degree}.
Let $W$ be the set of all support vertices. 
In order to prove ${\rm density}_x(W)  \geq \frac{9}{16}$, we need to show $|\hat E| = |\hat E[W]| \geq \frac{9}{8} \cdot (|W| - 1)$.
For this we use a token distribution argument.
Each vertex $v\in W$ initially receives $|\delta_{\hat E}(v)|$ tokens, i.e., one token per incident edge in $\hat G$.
The total number of tokens is therefore exactly $2|\hat E|$.

Each high-degree vertex now keeps $2+\frac{|\delta_{\hat E}(v)|-2}{|\delta_{\hat E}(v)|+1}$ tokens for itself, and distributes $\frac{|\delta_{\hat E}(v)|-2}{|\delta_{\hat E}(v)|+1} $ tokens to each one of its neighbors in $\hat G$.
Because  $|\delta_{\hat E}(v)| \geq 3$, each high-degree vertex keeps at least $2 + \frac{1}{4}$ tokens for itself and distributes at least $\frac{1}{4}$ tokens to each of its neighbors.
By Lemma~\ref{lem:dense_half_integral}, each low-degree vertex initially received two tokens.
It keeps these $2$ tokens and receives at least $\frac{1}{4}$ from a high-degree neighbor by \eqref{item:high-degree}.
Thus, after the redistribution of tokens each vertex in $W$ has at least $2 + \frac{1}{4}$ tokens.
Because the total number of tokens is $2 |\hat E|$, we conclude $2|\hat E|\geq (2+\frac{1}{4})|W|$ and therefore $|\hat E|  \geq \frac{9}{8} \cdot (|W| - 1)$.
\end{proof}

Having shown Lemma~\ref{lem:bound_mst_cost} and Lemma~\ref{lem:dense_half_integral}, we can prove that Algorithm~\ref{algorithm} returns a Steiner Forest solution of cost at most $\gapval \cdot c(x)$ whenever $(x,z)$ is half-integral.
This will prove Theorem~\ref{thm:half-integral}, which we restate here.

\halfintegral*

\begin{proof}
Consider Algorithm~\ref{algorithm} applied to a half-integral LP solution $(x,z)$.
By Lemma~\ref{lem:runtime_and_halfintegrality}, the running time is polynomial and in every recursive call of the algorithm, the given LP solution is half-integral.
We now prove that the algorithm returns a Steiner Forest solution of cost at most $\gapval \cdot c(x)$.

The modifications of the solution $(x,z)$ turning it into a well structured solution can only decrease the cost $c(x)$.
Let $(x_W,z_W)$ denote the LP solution that arises by contraction of $W$, i.e., the LP solution which we round to the edge set $F_W$ by calling the algorithms recursively.
Because the number of vertices decreases by at least one in every recursive call, we may assume by induction on the number of vertices, that $c(F_W) \leq \gapval \cdot c(x_W)$.
Moreover, by Lemma~\ref{lem:bound_mst_cost}, we have ${\rm mst}(W) \leq \frac{c(x|_{E[W]})}{{\rm density}_x(W)}$ and thus by Lemma~\ref{lem:dense_half_integral},
${\rm mst}(W) \leq \gapval \cdot c(x|_{E[W]})$.
Using $c(x_W) + (x|_{E[W]}) \leq c(x)$, this implies the claimed upper bound on the cost of the returned edge set $F$.

Finally, we observe that the returned edge set $F$ is a feasible solution to the given Steiner Forest instance.
Every pair $P\in \mathcal{P}$ with both elements in $W$ is connected by ${\rm MST}(W)$.
Moreover, for every pair $\{v,w\}\in \mathcal{P}$ with exactly one endpoint $w$ in $W$, the other endpoint $v$ is connected to some vertex $w'\in W$ by $F_W$ and $w$ is connected to $w$ by ${\rm MST}(W)$.
For every pair $\{u,v\}\in \mathcal{P}$ with neither endpoint in $W$, the set $F_W$ contains the edges of a $u$-$v$ path in the graph $\overline{G}/W$ arising by contraction of $W$.
Therefore, the union of $F_W$ and the tree ${\rm MST}(W)$ connecting all vertices in $W$ contains the edge set of a $u$-$v$ path in $\overline{G}$.
\end{proof}

\section{Steiner tree instances}\label{sec:steiner-tree-case}

In this section we prove Theorem~\ref{thm:steiner}, which we restate here for convenience.

\steiner*

We fix a root vertex $r_0\in R$. To prove Theorem~\ref{thm:steiner} we consider a solution $(x,z)$ to \ref{eq:forest-bcr} and construct a feasible solution to \ref{eq:bcr} with root $r_0$ for the Steiner tree instance $(G,c,R)$  with the same value.

We fix a spanning tree of the unique nontrivial connected component of the demand graph and orient it away from  the root $r_0$. 
Let $(R,A)$ be the resulting arborescence with root $r_0$.
Number the vertices in $R$ as $r_0, r_1, \dots, r_{|R|-1}$ in a topological order of $(R,A)$, i.e., such that $i < j$ for every arc $(r_i, r_j)\in A$.

In the following, we will first consider every vertex $w\in V$ separately. 
For each $w\in V$, we will consider the vector $x^w \in \mathbb{R}^{\overrightarrow{E}}$ and transform it into a vector $\overline{x}^w$ by reorienting some carefully chosen part of $x^w$.
Finally, we will prove that the vector $\tilde{x} \in \mathbb{R}^{\overrightarrow{E}}$, given by $\tilde{x}_e \coloneqq \sum_{w\in V} \overline{x}^w_e $ is a feasible solution to \ref{eq:bcr} for the Steiner Tree instance $(G,c,R)$ with root $r_0$.

We fix a vertex $w\in V$ and now describe how we construct the vector $\overline{x}^w$.
Ideally, we would like to reorient some part of $x^w$ to change the root from $w$ to $r_0$.
This strategy is feasible if an $r_0$-$w$ flow with edge capacities $x^w$ of value $\max \{ z^w_P : P \in \mathcal{P}\}$ exists, in which case we could reorient this flow $f$ by setting 
$
  \overline{x}^w_{(u,v)} \ \coloneqq\  x^w_{(u,v)} - f_{(u,v)} + f_{(v,u)}
$
for each edge $(u,v)\in \overrightarrow{E}$.
However, such a flow $f$ does not always exist and we will instead not necessarily send all the flow $f$ from $r_0$, but send it from different vertices $r_i$ -- aiming at sending as much flow as possible from vertices with small index $i$. 

More precisely, we  proceed as follows.
For each $i \in \{0, 1, \dots, |R|-1\}$, let 
\[
 \lambda_i \ \coloneqq\ \max \bigl\{ z^w_P : P \in \mathcal{P}\text{ with } P\cap\{r_0, r_1,  \dots, r_i\} \neq \emptyset \bigr\}.
\]
Moreover, let $\mu_0 = \lambda_0$ and  let $\mu_i  \coloneqq \lambda_i - \lambda_{i-1} \ge 0$ for $i \in \{1,\dots, |R|-1\}$.
Then there exists a flow $f$ that sends $\mu_i$-units to $w$ from all $r_i \in R$ and obeys edge capacities $x^w$. 

\begin{lemma}\label{lem:flow_steiner}
There exists a vector $f \in \mathbb{R}^{\overrightarrow{E}}$ with
\begin{align*}
f(\delta^+(r_i)) - f(\delta^-(r_i))\ =&\ \mu_i  & &\text{ for each }r_i\in R\setminus \{w\} \\
f(\delta^+(v)) - f(\delta^-(v))\ =&\  0 & &\text{ for each }v\in V\setminus R \\
f_e \ \le&\ x^w_e & &\text{ for each edge }e\in \overrightarrow{E}.
\end{align*}
\end{lemma}
\begin{proof}
Consider a set $U\subseteq V\setminus \{w\}$ and let $i_{\max} \coloneqq \max \{ i : r_i \in U\}$. 
Then 
\[
\sum _{i: r_i \in U} \mu_i \ \le\ \sum_{i=1}^{i_{\max}} \mu_i \ =\ \lambda_{i_{\max}}.
\]
Moreover, because $r_{i_{\max}} \in U$, there is a pair $P\in \mathcal{P}$ with $P\cap U \neq \emptyset$ and $z_P = \lambda_{i_{\max}}$, which implies $x^w(\delta^+(U)) \ge \lambda_{i_{\max}}$ by the constraints of \ref{eq:forest-bcr}.
This shows that the desired flow $f$ exists.
\end{proof}

Next we obtain $\overline{x}^w$ by reorienting this flow: for an edge $(u,v)\in \overrightarrow{E}$, we define
\[
  \overline{x}^w_{(u,v)} \ \coloneqq\  x^w_{(u,v)} - f_{(u,v)} + f_{(v,u)}.
\]
Because the flow $f$ obeys the edge capacities $x^w$, the vector $\overline{x}^w$ has only nonnegative entries.

\begin{lemma}\label{lem:feasible_for_bcr}
The vector $\tilde{x} \in \mathbb{R}^{\overrightarrow{E}}$, given by $\tilde{x}_e \coloneqq \sum_{w\in V} \overline{x}^w_e $ is a feasible solution to \ref{eq:bcr} for the Steiner Tree instance $(G,c,R)$ with root $r_0$.
\end{lemma}

Before proving the lemma, let us first observe that it implies Theorem~\ref{thm:steiner}.
\begin{proof}[Proof of Theorem \ref{thm:steiner}]
The theorem follows from Lemma \ref{lem:feasible_for_bcr} by using that the value $\sum_{w\in V} \sum_{e \in \overrightarrow{E}} c(e) \cdot x^w_e$ of the solution $(x,z)$ to \ref{eq:forest-bcr} is the same as the value $\sum_{e\in \overrightarrow{E}} c(e) \cdot \tilde{x}_e$ of the solution $\tilde{x}$ to \ref{eq:bcr}.   
\end{proof}

Recall that the arborescence $(R,A)$ arose from a spanning tree of the nontrivial component of the demand graph by orienting it away from the root $r_0$.
To prove Lemma~\ref{lem:feasible_for_bcr}, we first show the following:
\begin{lemma}\label{lem_steiner_cut_value_sufficient}
Let $w\in V$ and let $U\subseteq V\setminus \{r_0\}$ with $R\cap U \neq \emptyset$.
Let $j\coloneqq \min \{ i : r_i \in U\}$, let $(r_i, r_j) \in A$ be the incoming arc of $r_j$ in $A$, and let $P=\{r_i,r_j\}$.
Then $\overline{x}^w(\delta^+(U)) \ge z^w_P$.
\end{lemma}

\begin{proof}
We distinguish two cases. First, consider the case $w\in U$.
Then we have 
\begin{align*}
\overline{x}^w(\delta^+(U))  \ =&\ x^w(\delta^+(U)) - f(\delta^+(U)) + f(\delta^-(U)) \\
 =&\ x^w(\delta^+(U))  + \sum_{k: r_k \notin U} \mu_k \\
 \ge&\ \sum_{k: r_k \notin U} \mu_k.
\end{align*}

For every $k$ with $r_k\in U$, the choice of $j$ implies $k \ge j > i$ and thus we have $\{r_0, r_1, \dots, r_i\} \cap U = \emptyset$. 
We conclude 
\[
\overline{x}^w(\delta^+(U))  \ \ge\ \sum_{k: r_k \notin U} \mu_k \ \ge\ \sum_{k=0}^i \mu_k \ =\ \lambda_i \ \ge\ z^w_P,
\]
where we used $r_i \in P$ and the definition of $\lambda_i$ in the last inequality.

Now consider the remaining case $w\notin U$.
Let 
\[
z^{\max} \ \coloneqq\ \max \bigl\{ z_P : P \in \mathcal{P} \text{ with }P \cap U \neq \emptyset\bigr\}.
\]
Then the constraints of \ref{eq:forest-bcr} imply $x^w(\delta^+(U)) \geq z^{\max}$.
Thus, we have
\begin{align*}
\overline{x}^w(\delta^+(U))  \ =&\ x^w(\delta^+(U)) - f(\delta^+(U)) + f(\delta^-(U))  \\
=&\ x^w(\delta^+(U))  - \sum_{k: r_k \in U} \mu_k \\
\ge&\ z^{\max} - \sum_{k: r_k\in U} \mu_k.
\end{align*}
Using again that the choice of $j$ and $i$ implies $\{r_0, r_1, \dots, r_i\} \cap U = \emptyset$, we get $\sum_{k: r_k \in U} \mu_k \le z^{\max} - \sum_{k=0}^i \mu_k = z^{\max} - \lambda_i$.
Therefore,
\[
\overline{x}^w(\delta^+(U)) \ \ge\ z^{\max} - \sum_{k: r_k\in U} \mu_k \ \ge\ z^{\max} -  (z^{\max} - \lambda_i) \ =\ \lambda_i \ \ge\ z^w_P,
\]
where we again used $r_i \in P$ and the definition of $\lambda_i$ in the last inequality.
\end{proof}

We are now ready to complete the proof of Lemma~\ref{lem:feasible_for_bcr}.

\begin{proof}[Proof of Lemma~\ref{lem:feasible_for_bcr}]
First, we observe that $\overline{x}^w_e \ge 0$ for every vertex $w$ and every edge $e\in \overrightarrow{E}$ and hence $\tilde{x}_e \ge 0$ for every $e\in \overrightarrow{E}$.

Next, we consider a set $U\subseteq V\setminus \{r_0\}$ with $R\cap U \neq \emptyset$  and prove $\tilde{x}(\delta^+(U)) \ge 1$.
We let $j\coloneqq \min \{ i : r_i \in U\}$.
Let $(r_i, r_j) \in A$ be the incoming arc of $r_j$ in the arborescence $(R,A)$; then $i < j$.
By the definition of the demand graph, the pair $P\coloneqq \{r_i, r_j\}$ is contained in $\mathcal{P}$. 
From Lemma~\ref{lem_steiner_cut_value_sufficient} we get $\tilde{x}(\delta^+(U)) =  \sum_{w\in V} \overline{x}^w(\delta^+(U))  \geq 1$, because $\sum_{w\in V} z^w_P = 1$ by the constraints of \ref{eq:forest-bcr}. Therefore, the vector $\tilde{x}$ is a feasible solution to \ref{eq:bcr}.
\end{proof}

\section{Lower bound on the integrality gap}
\label{sec:lowerBound}

In this section we prove Theorem \ref{thm:lower_bound}, which we restate here for convenience.
\integralityGap*

\begin{proof}
We consider for every $q\in \mathbb{Z}_{>0}$, the following instance of Steiner Forest.
The graph $G$ has vertices $s_i$, $v_i$, and $t_i$ for $i\in\{1,\dots,q\}$.
The edges of $G$ consist of all edges between vertices $s_i$ and $v_j$, and all edges between vertices $v_i$ and $t_j$, i.e., we have
\[
 E \ \coloneqq\ \big\{ \{s_i, v_j\} :  i,j \in \{1,\dots,q\}  \big\} \cup \big\{ \{v_i, t_j\} :  i,j \in \{1,\dots,q\}  \big\}.
\]
See Figure~\ref{fig:integrality_gap}.
The set $\mathcal{P}$ contains the pairs $\{s_i,t_i\}$ for $i\in \{1,\dots,q\}$ and the pairs $\{v_i, v_{i+1}\}$ for $i\in \{1,\dots, q-1\}$.
The cost of every edge is $1$.

\begin{figure}
\begin{center}
\begin{tikzpicture}[xscale=0.9, yscale=1.2]

\tikzset{
terminal/.style={
draw=black,thick, circle,minimum size=1.3em, inner sep=0.2pt, outer sep=0.5pt}
}

\useasboundingbox (-5.9,-0.4) rectangle (9.9,3.6);

\begin{scope}[shift={(-5.5,0)}]

\begin{scope}[every node/.style={terminal}]
\node (s1) at (0,3) {$s_1$};
\node (s2) at (0,2) {$s_2$};
\node (s3) at (0,1) {$s_3$};
\node (s4) at (0,0) {$s_4$};
\node (v1) at (2,3) {$v_1$};
\node (v2) at (2,2) {$v_2$};
\node (v3) at (2,1) {$v_3$};
\node (v4) at (2,0) {$v_4$};
\node (t1) at (4,3) {$t_1$};
\node (t2) at (4,2) {$t_2$};
\node (t3) at (4,1) {$t_3$};
\node (t4) at (4,0) {$t_4$};
\end{scope}

\begin{scope}[very thick]
\draw (s1) -- (v1);
\draw (v1) -- (t1);
\draw (s1) -- (v2);
\draw (v2) -- (t1);
\draw (s1) -- (v3);
\draw (v3) -- (t1);
\draw (s1) -- (v4);
\draw (v4) -- (t1);
\draw (s2) -- (v1);
\draw (v1) -- (t2);
\draw (s2) -- (v2);
\draw (v2) -- (t2);
\draw (s2) -- (v3);
\draw (v3) -- (t2);
\draw (s2) -- (v4);
\draw (v4) -- (t2);
\draw (s3) -- (v1);
\draw (v1) -- (t3);
\draw (s3) -- (v2);
\draw (v2) -- (t3);
\draw (s3) -- (v3);
\draw (v3) -- (t3);
\draw (s3) -- (v4);
\draw (v4) -- (t3);
\draw (s4) -- (v1);
\draw (v1) -- (t4);
\draw (s4) -- (v2);
\draw (v2) -- (t4);
\draw (s4) -- (v3);
\draw (v3) -- (t4);
\draw (s4) -- (v4);
\draw (v4) -- (t4);
\end{scope}

\end{scope}

\begin{scope}

\begin{scope}[every node/.style={terminal}]
\node (s1) at (0,3) {$s_1$};
\node (s2) at (0,2) {$s_2$};
\node (s3) at (0,1) {$s_3$};
\node (s4) at (0,0) {$s_4$};
\node (v1) at (2,3) {$v_1$};
\node (v2) at (2,2) {$v_2$};
\node (v3) at (2,1) {$v_3$};
\node (v4) at (2,0) {$v_4$};
\node (t1) at (4,3) {$t_1$};
\node (t2) at (4,2) {$t_2$};
\node (t3) at (4,1) {$t_3$};
\node (t4) at (4,0) {$t_4$};
\end{scope}

\begin{scope}[ultra thick, densely dotted, cyan!80!black]
\draw[bend left=20] (s1) to (t1);
\draw[bend left=20] (s2) to (t2);
\draw[bend left=20] (s3) to (t3);
\draw[bend left=20] (s4) to (t4);
\end{scope}
\begin{scope}[ultra thick, densely dotted, violet]
\draw  (v1) to (v2);
\draw (v2) to (v3);
\draw (v3)  to (v4);
\end{scope}

\end{scope}

\begin{scope}[shift={(5.5,0)}]

\begin{scope}[every node/.style={terminal}]
\node (s1) at (0,3) {$s_1$};
\node (s2) at (0,2) {$s_2$};
\node (s3) at (0,1) {$s_3$};
\node (s4) at (0,0) {$s_4$};
\node (v1) at (2,3) {$v_1$};
\node (v2) at (2,2) {$v_2$};
\node (v3) at (2,1) {$v_3$};
\node (v4) at (2,0) {$v_4$};
\node[fill=purple, fill opacity=0.3, text=black, text opacity=1] (t1) at (4,3) {$t_1$};
\node[fill=green!60!black, fill opacity=0.3, text=black, text opacity=1] (t2) at (4,2) {$t_2$};
\node[fill=orange, fill opacity=0.3, text=black, text opacity=1] (t3) at (4,1) {$t_3$};
\node[fill=blue, fill opacity=0.3, text=black, text opacity=1] (t4) at (4,0) {$t_4$};
\end{scope}

\begin{scope}[very thick, ->, >=latex]

\begin{scope}[purple]
\draw (s1) -- (v1);
\draw (v1) -- (t1);
\draw (s1) -- (v2);
\draw (v2) -- (t1);
\draw (s1) -- (v3);
\draw (v3) -- (t1);
\draw (s1) -- (v4);
\draw (v4) -- (t1);
\end{scope}
\begin{scope}[green!60!black]
\draw (s2) -- (v1);
\draw (v1) -- (t2);
\draw (s2) -- (v2);
\draw (v2) -- (t2);
\draw (s2) -- (v3);
\draw (v3) -- (t2);
\draw (s2) -- (v4);
\draw (v4) -- (t2);
\end{scope}
\begin{scope}[orange]
\draw (s3) -- (v1);
\draw (v1) -- (t3);
\draw (s3) -- (v2);
\draw (v2) -- (t3);
\draw (s3) -- (v3);
\draw (v3) -- (t3);
\draw (s3) -- (v4);
\draw (v4) -- (t3);
\end{scope}
\begin{scope}[blue]
\draw (s4) -- (v1);
\draw (v1) -- (t4);
\draw (s4) -- (v2);
\draw (v2) -- (t4);
\draw (s4) -- (v3);
\draw (v3) -- (t4);
\draw (s4) -- (v4);
\draw (v4) -- (t4);
\end{scope}
\end{scope}

\end{scope}

\end{tikzpicture}
\end{center}
\caption{\label{fig:integrality_gap}
The instance and LP solution we use to establish an integrality gap lower bound for $q=4$.
The figure shows the graph $G=(V,E)$ (left), the demand graph of the instance (middle), and an illustration of the values of $x_e^r$ in the LP solution (right).
For an edge $e\in \overrightarrow{E}$ we have  $x^r_e = \frac{1}{q} = \frac{1}{4}$ if $e$ is drawn in the same color as the vertex $r$. 
All other variables $x^r_e$ are zero. In particular, for every non-colored vertex $r$, we have $x^r_e=0$ for all $e\in \overrightarrow{E}$.
}
\end{figure}
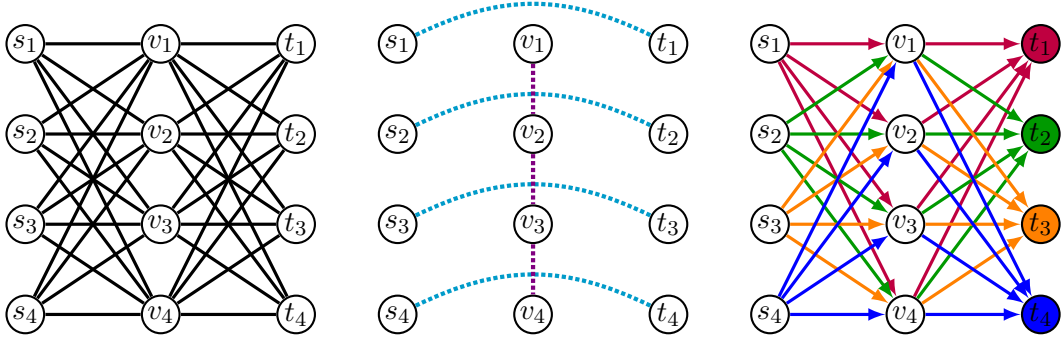

In any feasible solution $F$ of this instance, the vertices $v_1,\dots, v_q$ must belong to the same connected component of $(V,F)$.
Moreover, every path in $G$ from $s_i$ to $t_i$ contains some vertex $v_j$ and thus the vertices $s_i$ and $t_i$ must also belong to this connected component.
We conclude that $(V,F)$ has only a single connected component and is therefore a spanning tree of cost  $|V|-1 = 3q - 1$.

We now construct a feasible LP solution of cost $2q$.
For $i\in\{1,\dots,q\}$, we set
\[
z^r_{\{s_i,t_i\}}  \coloneqq 
\begin{cases}
1 &\text{ if } r= t_i \\
0 &\text{ otherwise}
\end{cases}
\]
and for $i\in\{1,\dots,q-1\}$, we set
\[
z^r_{\{v_i,v_{i+1}\}}  \coloneqq 
\begin{cases}
\frac{1}{q} &\text{ if } r= t_j \text{ for }j\in\{1,\dots,q\} \\
0 &\text{ otherwise.}
\end{cases}
\]
For $i,j\in \{1,\dots,q\}$, we set $x^{t_i}_{(s_i,v_j)} = x^{t_i}_{(v_j,t_i)} = \frac{1}{q}$. 
All other variables $x^r_e$ with $r\in V$ and $e \in \overrightarrow{E}$ are $0$.
This defines a feasible solution to \ref{eq:forest-bcr} of value $2q$.
Therefore, the integrality gap of the instance is at least $\frac{3q-1}{2q}$, which becomes arbitrarily close to $\frac{3}{2}$ for $q$ sufficiently large. 
\end{proof}

\section{Different representations can lead to different LP values }\label{sec:different_representations}

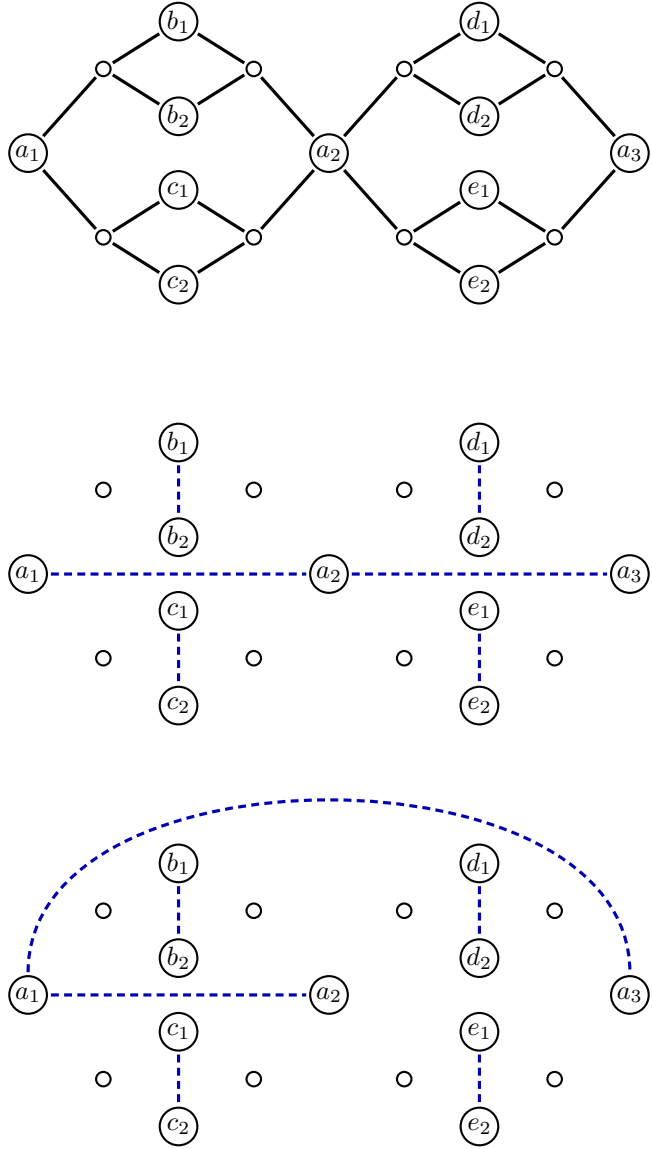
\begin{figure}
\begin{center}
\begin{tikzpicture}[yscale=0.7]

\tikzset{terminal/.style={
draw=black,thick, circle,minimum size=1.3em, inner sep=0.5pt, outer sep=1.5pt}
}

\tikzset{steiner/.style={
draw=black,thick, circle,minimum size=0.5em, inner sep=0.5pt, outer sep=1.5pt}
}

\begin{scope}[rotate=90,  xscale=-1]

\begin{scope}[every node/.style={terminal}]
\node (a1) at (0,4) {\small $a_1$};
\node (a2) at (0,0) {\small $a_2$};
\node (a3) at (0,-4) {\small $a_3$};
\node (b1) at (-2.5, 2) {\small $b_1$};
\node (b2) at (-0.7, 2) {\small $b_2$};
\node (c1) at (0.7, 2) {\small $c_1$};
\node (c2) at (2.5, 2) {\small $c_2$};
\node (d1) at (-2.5, -2) {\small $d_1$};
\node (d2) at (-0.7, -2) {\small $d_2$};
\node (e1) at (0.7, -2) {\small $e_1$};
\node (e2) at (2.5, -2) {\small $e_2$};
\end{scope}

\begin{scope}[every node/.style={steiner}]
\node (s1) at (-1.6, 3) {};
\node (s2) at (1.6, 3) {};
\node (s3) at (-1.6, 1) {};
\node (s4) at (1.6, 1) {};
\node (s5) at (-1.6, -1) {};
\node (s6) at (1.6, -1) {};
\node (s7) at (-1.6, -3) {};
\node (s8) at (1.6, -3) {};
\end{scope}

\begin{scope}[very thick]
\draw (a1) -- (s1);
\draw (a1) -- (s2);
\draw (s1) -- (b1);
\draw (s1) -- (b2);
\draw (s2) -- (c1);
\draw (s2) -- (c2);

\draw (s3) -- (b1);
\draw (s3) -- (b2);
\draw (s4) -- (c1);
\draw (s4) -- (c2);
\draw (a2) -- (s3);
\draw (a2) -- (s4);

\draw (a2) -- (s5);
\draw (a2) -- (s6);
\draw (s5) -- (d1);
\draw (s5) -- (d2);
\draw (s6) -- (e1);
\draw (s6) -- (e2);

\draw (s7) -- (d1);
\draw (s7) -- (d2);
\draw (s8) -- (e1);
\draw (s8) -- (e2);
\draw (a3) -- (s7);
\draw (a3) -- (s8);
\end{scope}

\end{scope}

\begin{scope}[shift={(0,-8)}, rotate=90,  xscale=-1]

\begin{scope}[every node/.style={terminal}]
\node (a1) at (0,4) {\small $a_1$};
\node (a2) at (0,0) {\small $a_2$};
\node (a3) at (0,-4) {\small $a_3$};
\node (b1) at (-2.5, 2) {\small $b_1$};
\node (b2) at (-0.7, 2) {\small $b_2$};
\node (c1) at (0.7, 2) {\small $c_1$};
\node (c2) at (2.5, 2) {\small $c_2$};
\node (d1) at (-2.5, -2) {\small $d_1$};
\node (d2) at (-0.7, -2) {\small $d_2$};
\node (e1) at (0.7, -2) {\small $e_1$};
\node (e2) at (2.5, -2) {\small $e_2$};
\end{scope}

\begin{scope}[every node/.style={steiner}]
\node (s1) at (-1.6, 3) {};
\node (s2) at (1.6, 3) {};
\node (s3) at (-1.6, 1) {};
\node (s4) at (1.6, 1) {};
\node (s5) at (-1.6, -1) {};
\node (s6) at (1.6, -1) {};
\node (s7) at (-1.6, -3) {};
\node (s8) at (1.6, -3) {};
\end{scope}

\begin{scope}[densely dashed, very thick, blue!70!black]
\draw (a1)-- (a2);
\draw (a2) -- (a3);
\draw  (b1) -- (b2);
\draw  (c1) -- (c2);
\draw  (d1) -- (d2);
\draw  (e1) -- (e2);
\end{scope}

\end{scope}

\begin{scope}[shift={(0, -16)}, rotate=90,  xscale=-1]

\begin{scope}[every node/.style={terminal}]
\node (a1) at (0,4) {\small $a_1$};
\node (a2) at (0,0) {\small $a_2$};
\node (a3) at (0,-4) {\small $a_3$};
\node (b1) at (-2.5, 2) {\small $b_1$};
\node (b2) at (-0.7, 2) {\small $b_2$};
\node (c1) at (0.7, 2) {\small $c_1$};
\node (c2) at (2.5, 2) {\small $c_2$};
\node (d1) at (-2.5, -2) {\small $d_1$};
\node (d2) at (-0.7, -2) {\small $d_2$};
\node (e1) at (0.7, -2) {\small $e_1$};
\node (e2) at (2.5, -2) {\small $e_2$};
\end{scope}

\begin{scope}[every node/.style={steiner}]
\node (s1) at (-1.6, 3) {};
\node (s2) at (1.6, 3) {};
\node (s3) at (-1.6, 1) {};
\node (s4) at (1.6, 1) {};
\node (s5) at (-1.6, -1) {};
\node (s6) at (1.6, -1) {};
\node (s7) at (-1.6, -3) {};
\node (s8) at (1.6, -3) {};
\end{scope}

\begin{scope}[densely dashed, very thick, blue!70!black]
\draw (a1)-- (a2);
\draw (a1)to[bend right=90, looseness =1.4] (a3);
\draw  (b1) -- (b2);
\draw  (c1) -- (c2);
\draw  (d1) -- (d2);
\draw  (e1) -- (e2);
\end{scope}

\end{scope}

\end{tikzpicture}
\end{center}
\caption{\label{fig:different_representations}
Illustration of the graph $G=(V,E)$ (top), the demand graph for the pairs $\mathcal{P}_1$ (middle),  and the demand graph for a different representation $\mathcal{P}_2$ of $\mathcal{P}_1$ (bottom).
All edges have cost one.
}
\end{figure}

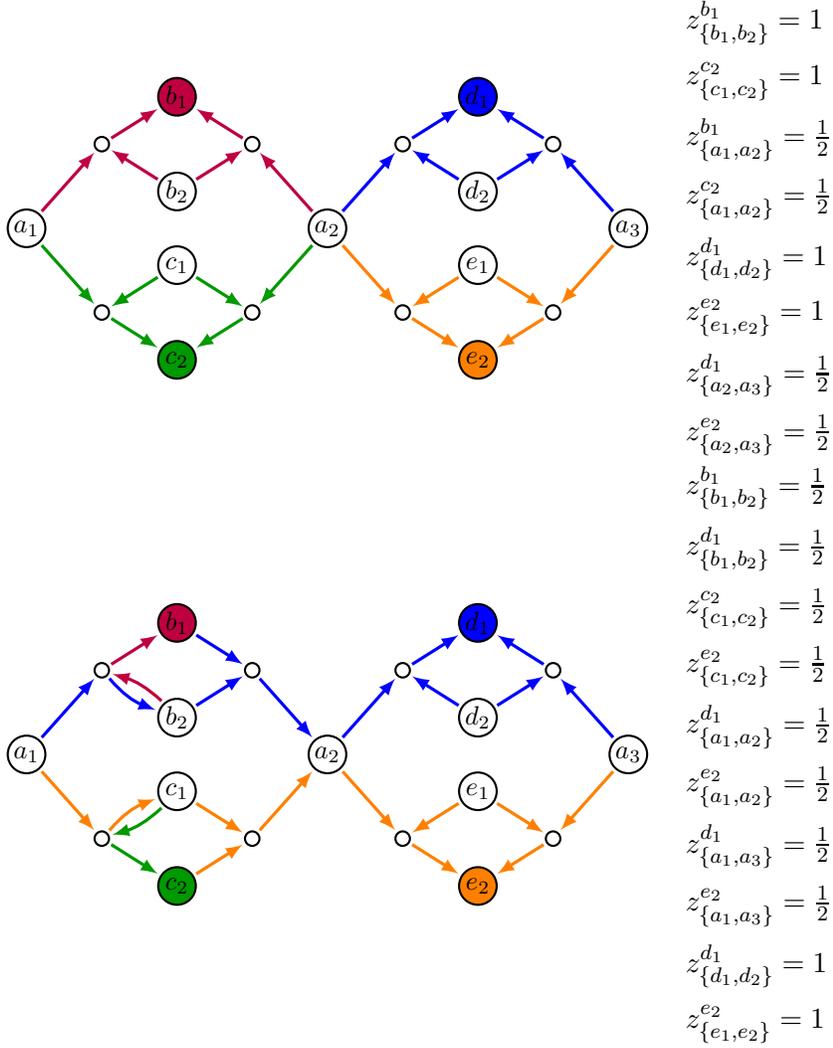
\begin{figure}[p]
\begin{center}
\begin{tikzpicture}[yscale=0.7]

\tikzset{terminal/.style={
draw=black,thick, circle,minimum size=1.3em, inner sep=0.5pt, outer sep=1.5pt}
}

\tikzset{steiner/.style={
draw=black,thick, circle,minimum size=0.5em, inner sep=0.5pt, outer sep=1.5pt}
}

\begin{scope}[rotate=90,  xscale=-1]

\begin{scope}[every node/.style={terminal}]
\node (a1) at (0,4) {\small $a_1$};
\node (a2) at (0,0) {\small $a_2$};
\node (a3) at (0,-4) {\small $a_3$};
\node[fill=purple, fill opacity=0.3, text=black, text opacity=1] (b1) at (-2.5, 2) {\small $b_1$};
\node (b2) at (-0.7, 2) {\small $b_2$};
\node (c1) at (0.7, 2) {\small $c_1$};
\node[fill=green!60!black, fill opacity=0.3, text=black, text opacity=1] (c2) at (2.5, 2) {\small $c_2$};
\node[fill=blue, fill opacity=0.3, text=black, text opacity=1] (d1) at (-2.5, -2) {\small $d_1$};
\node (d2) at (-0.7, -2) {\small $d_2$};
\node (e1) at (0.7, -2) {\small $e_1$};
\node[fill=orange, fill opacity=0.3, text=black, text opacity=1] (e2) at (2.5, -2) {\small $e_2$};
\end{scope}

\begin{scope}[every node/.style={steiner}]
\node (s1) at (-1.6, 3) {};
\node (s2) at (1.6, 3) {};
\node (s3) at (-1.6, 1) {};
\node (s4) at (1.6, 1) {};
\node (s5) at (-1.6, -1) {};
\node (s6) at (1.6, -1) {};
\node (s7) at (-1.6, -3) {};
\node (s8) at (1.6, -3) {};
\end{scope}

\begin{scope}[very thick, ->, >=latex]

\begin{scope}[purple]
\draw (a1) -- (s1);
\draw (b2) -- (s1);
\draw (s1) -- (b1);
\draw (s3) -- (b1);
\draw (b2) -- (s3);
\draw (a2) -- (s3);
\end{scope}

\begin{scope}[green!60!black]
\draw (a1) -- (s2);
\draw (c1) -- (s2);
\draw (s2) -- (c2);
\draw (c1) -- (s4);
\draw (s4) -- (c2);
\draw (a2) -- (s4);
\end{scope}

\begin{scope}[blue]
\draw (a2) -- (s5);
\draw (d2) -- (s5);
\draw (s5) -- (d1);
\draw (s7) -- (d1);
\draw (d2) -- (s7);
\draw (a3) -- (s7);
\end{scope}

\begin{scope}[orange]
\draw (a2) -- (s6);
\draw (e1) -- (s6);
\draw (s6) -- (e2);
\draw (e1) -- (s8);
\draw (s8) -- (e2);
\draw (a3) -- (s8);
\end{scope}

\end{scope}

\end{scope}

\node () at (6,0) {
\begin{minipage}{0.15\textwidth}
$z^{b_1}_{\{b_1,b_2\}} = 1 $ \\[2mm]
$z^{c_2}_{\{c_1,c_2\}} = 1 $  \\[2mm]
$z^{b_1}_{\{a_1,a_2\}} = \frac{1}{2} $ \\[2mm]
$z^{c_2}_{\{a_1,a_2\}} = \frac{1}{2} $  \\[2mm]
$z^{d_1}_{\{d_1,d_2\}} = 1 $ \\[2mm]
$z^{e_2}_{\{e_1,e_2\}} = 1 $  \\[2mm]
$z^{d_1}_{\{a_2,a_3\}} = \frac{1}{2} $ \\[2mm]
$z^{e_2}_{\{a_2,a_3\}} = \frac{1}{2} $ 
\end{minipage}
};

\begin{scope}[shift={(0,-10)}, rotate=90, xscale=-1]

\begin{scope}[every node/.style={terminal}]
\node (a1) at (0,4) {\small $a_1$};
\node (a2) at (0,0) {\small $a_2$};
\node (a3) at (0,-4) {\small $a_3$};
\node[fill=purple, fill opacity=0.3, text=black, text opacity=1] (b1) at (-2.5, 2) {\small $b_1$};
\node (b2) at (-0.7, 2) {\small $b_2$};
\node (c1) at (0.7, 2) {\small $c_1$};
\node[fill=green!60!black, fill opacity=0.3, text=black, text opacity=1] (c2) at (2.5, 2) {\small $c_2$};
\node[fill=blue, fill opacity=0.3, text=black, text opacity=1] (d1) at (-2.5, -2) {\small $d_1$};
\node (d2) at (-0.7, -2) {\small $d_2$};
\node (e1) at (0.7, -2) {\small $e_1$};
\node[fill=orange, fill opacity=0.3, text=black, text opacity=1] (e2) at (2.5, -2) {\small $e_2$};
\end{scope}

\begin{scope}[every node/.style={steiner}]
\node (s1) at (-1.6, 3) {};
\node (s2) at (1.6, 3) {};
\node (s3) at (-1.6, 1) {};
\node (s4) at (1.6, 1) {};
\node (s5) at (-1.6, -1) {};
\node (s6) at (1.6, -1) {};
\node (s7) at (-1.6, -3) {};
\node (s8) at (1.6, -3) {};
\end{scope}

\begin{scope}[very thick, ->, >=latex]

\begin{scope}[purple]
\draw (b2) to[bend left=15] (s1);
\draw (s1) -- (b1);
\end{scope}

\begin{scope}[green!60!black]
\draw (c1) to[bend right=15] (s2);
\draw (s2) -- (c2);
\end{scope}

\begin{scope}[blue]
\draw (a1) -- (s1);
\draw (s1) to [bend left=15] (b2);
\draw (b2) -- (s3);
\draw (s3) -- (a2);
\draw (b1) -- (s3);

\draw (a2) -- (s5);
\draw (d2) -- (s5);
\draw (s5) -- (d1);
\draw (s7) -- (d1);
\draw (d2) -- (s7);
\draw (a3) -- (s7);
\end{scope}

\begin{scope}[orange]
\draw (a1) -- (s2);
\draw (s2) to [bend right=15] (c1);
\draw (c1) -- (s4);
\draw (s4) -- (a2);
\draw (c2) -- (s4);

\draw (a2) -- (s6);
\draw (e1) -- (s6);
\draw (s6) -- (e2);
\draw (e1) -- (s8);
\draw (s8) -- (e2);
\draw (a3) -- (s8);
\end{scope}

\end{scope}

\end{scope}

\node () at (6,-10) {
\begin{minipage}{0.15\textwidth}
$z^{b_1}_{\{b_1,b_2\}} = \frac{1}{2} $ \\[2mm]
$z^{d_1}_{\{b_1,b_2\}} = \frac{1}{2} $ \\[2mm]
$z^{c_2}_{\{c_1,c_2\}} = \frac{1}{2} $  \\[2mm]
$z^{e_2}_{\{c_1,c_2\}} = \frac{1}{2} $  \\[2mm]
$z^{d_1}_{\{a_1,a_2\}} = \frac{1}{2} $ \\[2mm]
$z^{e_2}_{\{a_1,a_2\}} = \frac{1}{2} $  \\[2mm]
$z^{d_1}_{\{a_1,a_3\}} = \frac{1}{2} $ \\[2mm]
$z^{e_2}_{\{a_1,a_3\}} = \frac{1}{2} $ \\[2mm]
$z^{d_1}_{\{d_1,d_2\}} = 1 $ \\[2mm]
$z^{e_2}_{\{e_1,e_2\}} = 1 $  
\end{minipage}
};

\end{tikzpicture}
\end{center}
\caption{\label{fig:different_primal_solutions}
Optimal solutions to \ref{eq:forest-bcr} for the two different instances  $((V,E), c, \mathcal{P}_1)$ (top) and $((V,E), c, \mathcal{P}_2)$ (bottom).
For all red edges we have $x^{b_1}_e = \frac{1}{2}$, for all green edges we have $x^{c_2}_e = \frac{1}{2}$, for all blue edges we have $x^{d_1}_e = \frac{1}{2}$, and for all orange edges, we have $x^{e_2}_e = \frac{1}{2}$. 
All other variables $x^r_e$ are zero.
For the variables $z^r_P$, all positive values are shown.
The solution shown for the instance $((V,E), c, \mathcal{P}_1)$  has value $12$ and the solution shown for the instance $((V,E), c, \mathcal{P}_2)$ has value $13$.
}
\end{figure}

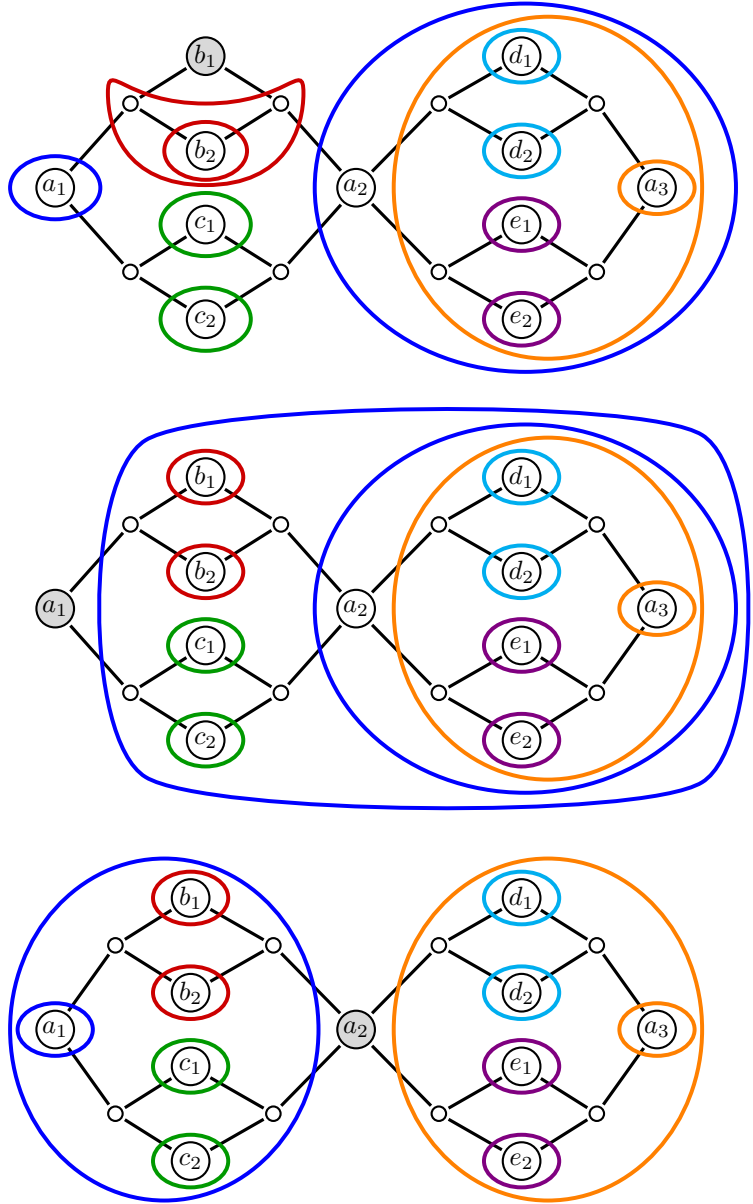
\begin{figure}[p]
\begin{center}
\begin{tikzpicture}[yscale=0.7]

\tikzset{terminal/.style={
draw=black,thick, circle,minimum size=1.3em, inner sep=0.5pt, outer sep=1.5pt}
}

\tikzset{steiner/.style={
draw=black,thick, circle,minimum size=0.5em, inner sep=0.5pt, outer sep=1.5pt}
}

\tikzset{dual/.style={line width=1.5pt}}

\begin{scope}[rotate=90,  xscale=-1]

\begin{scope}[every node/.style={terminal}]
\node (a1) at (0,4) {\small $a_1$};
\node (a2) at (0,0) {\small $a_2$};
\node (a3) at (0,-4) {\small $a_3$};
\node[fill=gray!30!white, text=black] (b1) at (-2.5, 2) {\small $b_1$};
\node (b2) at (-0.7, 2) {\small $b_2$};
\node (c1) at (0.7, 2) {\small $c_1$};
\node (c2) at (2.5, 2) {\small $c_2$};
\node (d1) at (-2.5, -2.2) {\small $d_1$};
\node (d2) at (-0.7, -2.2) {\small $d_2$};
\node (e1) at (0.7, -2.2) {\small $e_1$};
\node (e2) at (2.5, -2.2) {\small $e_2$};
\end{scope}

\begin{scope}[every node/.style={steiner}]
\node (s1) at (-1.6, 3) {};
\node (s2) at (1.6, 3) {};
\node (s3) at (-1.6, 1) {};
\node (s4) at (1.6, 1) {};
\node (s5) at (-1.6, -1.1) {};
\node (s6) at (1.6, -1.1) {};
\node (s7) at (-1.6, -3.2) {};
\node (s8) at (1.6, -3.2) {};
\end{scope}

\begin{scope}[very thick]
\draw (a1) -- (s1);
\draw (a1) -- (s2);
\draw (s1) -- (b1);
\draw (s1) -- (b2);
\draw (s2) -- (c1);
\draw (s2) -- (c2);

\draw (s3) -- (b1);
\draw (s3) -- (b2);
\draw (s4) -- (c1);
\draw (s4) -- (c2);
\draw (a2) -- (s3);
\draw (a2) -- (s4);

\draw (a2) -- (s5);
\draw (a2) -- (s6);
\draw (s5) -- (d1);
\draw (s5) -- (d2);
\draw (s6) -- (e1);
\draw (s6) -- (e2);

\draw (s7) -- (d1);
\draw (s7) -- (d2);
\draw (s8) -- (e1);
\draw (s8) -- (e2);
\draw (a3) -- (s7);
\draw (a3) -- (s8);
\end{scope}

\begin{scope}[dual]
\begin{scope}[blue]
\draw (a1) ellipse (0.6 and 0.6);
\draw (0,-2.25) ellipse (3.5 and 2.8);
\end{scope}
\begin{scope}[red!80!black]
\draw (b2) ellipse (0.55 and 0.55);
\draw [dual] plot [smooth cycle, tension=1.5] coordinates { (-0.05,2) (-1.8,3.3) (-1.6,2) (-1.8,0.7)};
\end{scope}
\begin{scope}[green!60!black]
\draw (c1) ellipse (0.6 and 0.6);
\draw (c2) ellipse (0.6 and 0.6);
\end{scope}
\begin{scope}[orange]
\draw (a3) ellipse (0.5 and 0.5);
\draw (0,-2.55) ellipse (3.25 and 2.05);
\end{scope}
\begin{scope}[cyan]
\draw (d1) ellipse (0.5 and 0.5);
\draw (d2) ellipse (0.5 and 0.5);
\end{scope}
\begin{scope}[violet]
\draw (e1) ellipse (0.5 and 0.5);
\draw (e2) ellipse (0.5 and 0.5);
\end{scope}
\end{scope}

\end{scope}

\begin{scope}[shift={(0,-8)}, rotate=90,  xscale=-1]

\begin{scope}[every node/.style={terminal}]
\node[fill=gray!30!white, text=black]  (a1) at (0,4) {\small $a_1$};
\node (a2) at (0,0) {\small $a_2$};
\node (a3) at (0,-4) {\small $a_3$};
\node (b1) at (-2.5, 2) {\small $b_1$};
\node (b2) at (-0.7, 2) {\small $b_2$};
\node (c1) at (0.7, 2) {\small $c_1$};
\node (c2) at (2.5, 2) {\small $c_2$};
\node (d1) at (-2.5, -2.2) {\small $d_1$};
\node (d2) at (-0.7, -2.2) {\small $d_2$};
\node (e1) at (0.7, -2.2) {\small $e_1$};
\node (e2) at (2.5, -2.2) {\small $e_2$};
\end{scope}

\begin{scope}[every node/.style={steiner}]
\node (s1) at (-1.6, 3) {};
\node (s2) at (1.6, 3) {};
\node (s3) at (-1.6, 1) {};
\node (s4) at (1.6, 1) {};
\node (s5) at (-1.6, -1.1) {};
\node (s6) at (1.6, -1.1) {};
\node (s7) at (-1.6, -3.2) {};
\node (s8) at (1.6, -3.2) {};
\end{scope}

\begin{scope}[very thick]
\draw (a1) -- (s1);
\draw (a1) -- (s2);
\draw (s1) -- (b1);
\draw (s1) -- (b2);
\draw (s2) -- (c1);
\draw (s2) -- (c2);

\draw (s3) -- (b1);
\draw (s3) -- (b2);
\draw (s4) -- (c1);
\draw (s4) -- (c2);
\draw (a2) -- (s3);
\draw (a2) -- (s4);

\draw (a2) -- (s5);
\draw (a2) -- (s6);
\draw (s5) -- (d1);
\draw (s5) -- (d2);
\draw (s6) -- (e1);
\draw (s6) -- (e2);

\draw (s7) -- (d1);
\draw (s7) -- (d2);
\draw (s8) -- (e1);
\draw (s8) -- (e2);
\draw (a3) -- (s7);
\draw (a3) -- (s8);
\end{scope}

\begin{scope}[dual]
\begin{scope}[blue]
\draw [dual] plot [smooth cycle, tension=0.4] coordinates { (-3.2,2.8) (3.2,2.8) (3.3,-4.6) (-3.3,-4.6)};
\draw (0,-2.25) ellipse (3.5 and 2.8);
\end{scope}
\begin{scope}[red!80!black]
\draw (b1) ellipse (0.5 and 0.5);
\draw (b2) ellipse (0.5 and 0.5);
\end{scope}
\begin{scope}[green!60!black]
\draw (c1) ellipse (0.5 and 0.5);
\draw (c2) ellipse (0.5 and 0.5);
\end{scope}
\begin{scope}[orange]
\draw (a3) ellipse (0.5 and 0.5);
\draw (0,-2.55) ellipse (3.25 and 2.05);
\end{scope}
\begin{scope}[cyan]
\draw (d1) ellipse (0.5 and 0.5);
\draw (d2) ellipse (0.5 and 0.5);
\end{scope}
\begin{scope}[violet]
\draw (e1) ellipse (0.5 and 0.5);
\draw (e2) ellipse (0.5 and 0.5);
\end{scope}
\end{scope}

\end{scope}

\begin{scope}[shift={(0,-16)}, rotate=90,  xscale=-1]

\begin{scope}[every node/.style={terminal}]
\node (a1) at (0,4) {\small $a_1$};
\node[fill=gray!30!white, text=black] (a2) at (0,0) {\small $a_2$};
\node (a3) at (0,-4) {\small $a_3$};
\node (b1) at (-2.5, 2.2) {\small $b_1$};
\node (b2) at (-0.7, 2.2) {\small $b_2$};
\node (c1) at (0.7, 2.2) {\small $c_1$};
\node (c2) at (2.5, 2.2) {\small $c_2$};
\node (d1) at (-2.5, -2.2) {\small $d_1$};
\node (d2) at (-0.7, -2.2) {\small $d_2$};
\node (e1) at (0.7, -2.2) {\small $e_1$};
\node (e2) at (2.5, -2.2) {\small $e_2$};
\end{scope}

\begin{scope}[every node/.style={steiner}]
\node (s1) at (-1.6, 3.2) {};
\node (s2) at (1.6, 3.2) {};
\node (s3) at (-1.6, 1.1) {};
\node (s4) at (1.6, 1.1) {};
\node (s5) at (-1.6, -1.1) {};
\node (s6) at (1.6, -1.1) {};
\node (s7) at (-1.6, -3.2) {};
\node (s8) at (1.6, -3.2) {};
\end{scope}

\begin{scope}[very thick]
\draw (a1) -- (s1);
\draw (a1) -- (s2);
\draw (s1) -- (b1);
\draw (s1) -- (b2);
\draw (s2) -- (c1);
\draw (s2) -- (c2);

\draw (s3) -- (b1);
\draw (s3) -- (b2);
\draw (s4) -- (c1);
\draw (s4) -- (c2);
\draw (a2) -- (s3);
\draw (a2) -- (s4);

\draw (a2) -- (s5);
\draw (a2) -- (s6);
\draw (s5) -- (d1);
\draw (s5) -- (d2);
\draw (s6) -- (e1);
\draw (s6) -- (e2);

\draw (s7) -- (d1);
\draw (s7) -- (d2);
\draw (s8) -- (e1);
\draw (s8) -- (e2);
\draw (a3) -- (s7);
\draw (a3) -- (s8);
\end{scope}

\begin{scope}[dual]
\begin{scope}[blue]
\draw (a1) ellipse (0.5 and 0.5);
\draw (0,2.55) ellipse (3.25 and 2.05);
\end{scope}
\begin{scope}[red!80!black]
\draw (b1) ellipse (0.5 and 0.5);
\draw (b2) ellipse (0.5 and 0.5);
\end{scope}
\begin{scope}[green!60!black]
\draw (c1) ellipse (0.5 and 0.5);
\draw (c2) ellipse (0.5 and 0.5);
\end{scope}
\begin{scope}[orange]
\draw (a3) ellipse (0.5 and 0.5);
\draw (0,-2.55) ellipse (3.25 and 2.05);
\end{scope}
\begin{scope}[cyan]
\draw (d1) ellipse (0.5 and 0.5);
\draw (d2) ellipse (0.5 and 0.5);
\end{scope}
\begin{scope}[violet]
\draw (e1) ellipse (0.5 and 0.5);
\draw (e2) ellipse (0.5 and 0.5);
\end{scope}
\end{scope}

\end{scope}

\end{tikzpicture}
\end{center}
\caption{\label{fig:dual_solution_1}
The picture illustrates the values $y^r_{U,P}$ for $r=b_1$ (top), $r=a_1$ (middle) and $r=a_2$ (bottom) for an optimal dual solution for  $((V,E), c, \mathcal{P}_1)$.
The colored sets indicate non-zero variables $y^r_{U,P}$, where the color of the drawn set $U$ indicates the pair $P$.
For every drawn set $U$, we have a variable $y^r_{U,P}$ with value $1$, where \textcolor{blue}{$P=\{a_1,a_2\}$} for darkblue sets, \textcolor{orange}{$P=\{a_2, a_3\}$} for orange sets,
\textcolor{red}{$P=\{b_1,b_2\}$} for red sets, \textcolor{green!60!black}{$P=\{c_1,c_2\}$} for green sets, \textcolor{cyan}{$P=\{d_1,d_2\}$} for lightblue sets, and \textcolor{violet}{$P=\{e_1,e_2\}$} for violet sets.
}
\end{figure}

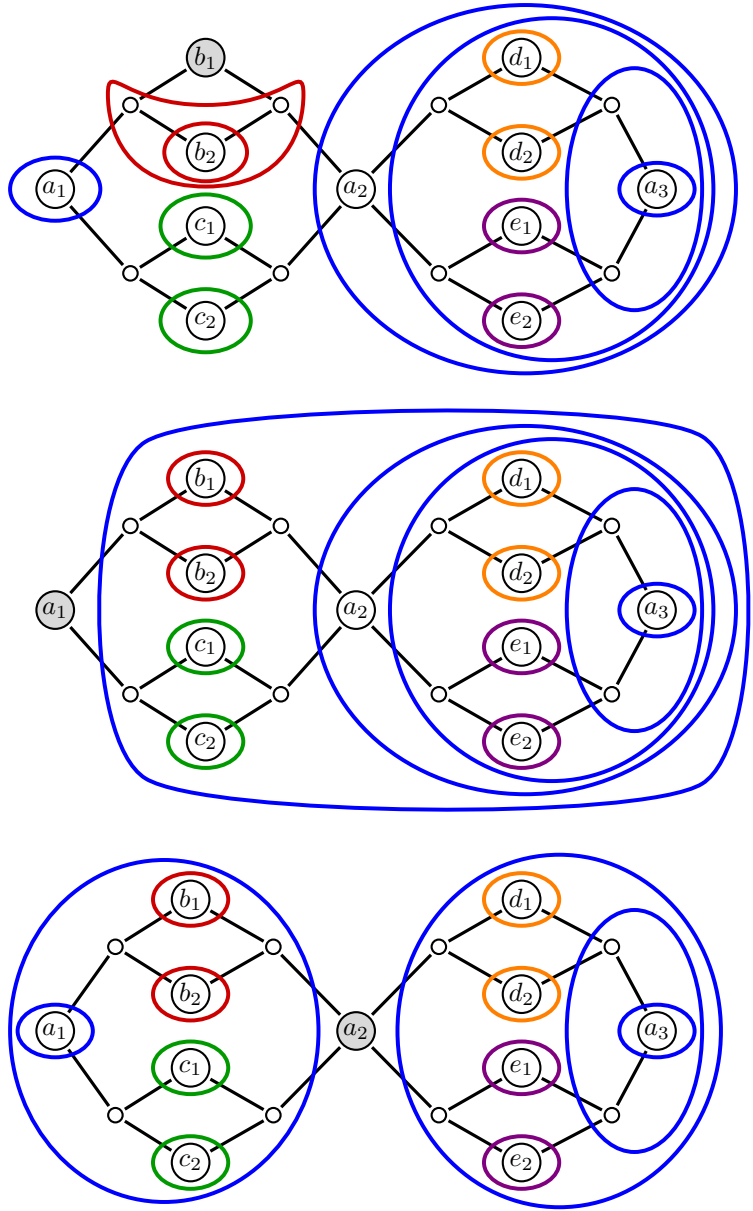
\begin{figure}[p]
\begin{center}
\begin{tikzpicture}[yscale=0.7]

\tikzset{terminal/.style={
draw=black,thick, circle,minimum size=1.3em, inner sep=0.5pt, outer sep=1.5pt}
}

\tikzset{steiner/.style={
draw=black,thick, circle,minimum size=0.5em, inner sep=0.5pt, outer sep=1.5pt}
}

\tikzset{dual/.style={line width=1.5pt}}

\begin{scope}[rotate=90, xscale=-1]

\begin{scope}[every node/.style={terminal}]
\node (a1) at (0,4) {\small $a_1$};
\node (a2) at (0,0) {\small $a_2$};
\node (a3) at (0,-4) {\small $a_3$};
\node[fill=gray!30!white, text=black] (b1) at (-2.5, 2) {\small $b_1$};
\node (b2) at (-0.7, 2) {\small $b_2$};
\node (c1) at (0.7, 2) {\small $c_1$};
\node (c2) at (2.5, 2) {\small $c_2$};
\node (d1) at (-2.5, -2.2) {\small $d_1$};
\node (d2) at (-0.7, -2.2) {\small $d_2$};
\node (e1) at (0.7, -2.2) {\small $e_1$};
\node (e2) at (2.5, -2.2) {\small $e_2$};
\end{scope}

\begin{scope}[every node/.style={steiner}]
\node (s1) at (-1.6, 3) {};
\node (s2) at (1.6, 3) {};
\node (s3) at (-1.6, 1) {};
\node (s4) at (1.6, 1) {};
\node (s5) at (-1.6, -1.1) {};
\node (s6) at (1.6, -1.1) {};
\node (s7) at (-1.6, -3.4) {};Lemma
\node (s8) at (1.6, -3.4) {};
\end{scope}

\begin{scope}[very thick]
\draw (a1) -- (s1);
\draw (a1) -- (s2);
\draw (s1) -- (b1);
\draw (s1) -- (b2);
\draw (s2) -- (c1);
\draw (s2) -- (c2);

\draw (s3) -- (b1);
\draw (s3) -- (b2);
\draw (s4) -- (c1);
\draw (s4) -- (c2);
\draw (a2) -- (s3);
\draw (a2) -- (s4);

\draw (a2) -- (s5);
\draw (a2) -- (s6);
\draw (s5) -- (d1);
\draw (s5) -- (d2);
\draw (s6) -- (e1);
\draw (s6) -- (e2);

\draw (s7) -- (d1);
\draw (s7) -- (d2);
\draw (s8) -- (e1);
\draw (s8) -- (e2);
\draw (a3) -- (s7);
\draw (a3) -- (s8);
\end{scope}

\begin{scope}[dual]
\begin{scope}[blue]
\draw (a1) ellipse (0.6 and 0.6);
\draw (0,-2.25) ellipse (3.5 and 2.8);
\draw (0,-3.7) ellipse (2.3 and 0.9);
\draw (a3) ellipse (0.5 and 0.5);
\draw (0,-2.6) ellipse (3.25 and 2.15);
\end{scope}
\begin{scope}[red!80!black]
\draw (b2) ellipse (0.55 and 0.55);
\draw [dual] plot [smooth cycle, tension=1.5] coordinates { (-0.05,2) (-1.8,3.3) (-1.6,2) (-1.8,0.7)};
\end{scope}
\begin{scope}[green!60!black]
\draw (c1) ellipse (0.6 and 0.6);
\draw (c2) ellipse (0.6 and 0.6);
\end{scope}
\begin{scope}[orange]
\draw (d1) ellipse (0.5 and 0.5);
\draw (d2) ellipse (0.5 and 0.5);
\end{scope}
\begin{scope}[violet]
\draw (e1) ellipse (0.5 and 0.5);
\draw (e2) ellipse (0.5 and 0.5);
\end{scope}
\end{scope}

\end{scope}

\begin{scope}[shift={(0,-8)}, rotate=90, xscale=-1]

\begin{scope}[every node/.style={terminal}]
\node[fill=gray!30!white, text=black]  (a1) at (0,4) {\small $a_1$};
\node (a2) at (0,0) {\small $a_2$};
\node (a3) at (0,-4) {\small $a_3$};
\node (b1) at (-2.5, 2) {\small $b_1$};
\node (b2) at (-0.7, 2) {\small $b_2$};
\node (c1) at (0.7, 2) {\small $c_1$};
\node (c2) at (2.5, 2) {\small $c_2$};
\node (d1) at (-2.5, -2.2) {\small $d_1$};
\node (d2) at (-0.7, -2.2) {\small $d_2$};
\node (e1) at (0.7, -2.2) {\small $e_1$};
\node (e2) at (2.5, -2.2) {\small $e_2$};
\end{scope}

\begin{scope}[every node/.style={steiner}]
\node (s1) at (-1.6, 3) {};
\node (s2) at (1.6, 3) {};
\node (s3) at (-1.6, 1) {};
\node (s4) at (1.6, 1) {};
\node (s5) at (-1.6, -1.1) {};
\node (s6) at (1.6, -1.1) {};
\node (s7) at (-1.6, -3.4) {};
\node (s8) at (1.6, -3.4) {};
\end{scope}

\begin{scope}[very thick]
\draw (a1) -- (s1);
\draw (a1) -- (s2);
\draw (s1) -- (b1);
\draw (s1) -- (b2);
\draw (s2) -- (c1);
\draw (s2) -- (c2);

\draw (s3) -- (b1);
\draw (s3) -- (b2);
\draw (s4) -- (c1);
\draw (s4) -- (c2);
\draw (a2) -- (s3);
\draw (a2) -- (s4);

\draw (a2) -- (s5);
\draw (a2) -- (s6);
\draw (s5) -- (d1);
\draw (s5) -- (d2);
\draw (s6) -- (e1);
\draw (s6) -- (e2);

\draw (s7) -- (d1);
\draw (s7) -- (d2);
\draw (s8) -- (e1);
\draw (s8) -- (e2);
\draw (a3) -- (s7);
\draw (a3) -- (s8);
\end{scope}

\begin{scope}[dual]
\begin{scope}[blue]
\draw [dual] plot [smooth cycle, tension=0.4] coordinates { (-3.2,2.8) (3.2,2.8) (3.3,-4.6) (-3.3,-4.6)};
\draw (0,-2.25) ellipse (3.5 and 2.8);
\draw (0,-3.7) ellipse (2.3 and 0.9);
\draw (a3) ellipse (0.5 and 0.5);
\draw (0,-2.6) ellipse (3.25 and 2.15);
\end{scope}
\begin{scope}[red!80!black]
\draw (b1) ellipse (0.5 and 0.5);
\draw (b2) ellipse (0.5 and 0.5);
\end{scope}
\begin{scope}[green!60!black]
\draw (c1) ellipse (0.5 and 0.5);
\draw (c2) ellipse (0.5 and 0.5);
\end{scope}
\begin{scope}[orange]
\draw (d1) ellipse (0.5 and 0.5);
\draw (d2) ellipse (0.5 and 0.5);
\end{scope}
\begin{scope}[violet]
\draw (e1) ellipse (0.5 and 0.5);
\draw (e2) ellipse (0.5 and 0.5);
\end{scope}
\end{scope}

\end{scope}

\begin{scope}[shift={(0,-16)}, rotate=90,  xscale=-1]

\begin{scope}[every node/.style={terminal}]
\node (a1) at (0,4) {\small $a_1$};
\node[fill=gray!30!white, text=black] (a2) at (0,0) {\small $a_2$};
\node (a3) at (0,-4) {\small $a_3$};
\node (b1) at (-2.5, 2.2) {\small $b_1$};
\node (b2) at (-0.7, 2.2) {\small $b_2$};
\node (c1) at (0.7, 2.2) {\small $c_1$};
\node (c2) at (2.5, 2.2) {\small $c_2$};
\node (d1) at (-2.5, -2.2) {\small $d_1$};
\node (d2) at (-0.7, -2.2) {\small $d_2$};
\node (e1) at (0.7, -2.2) {\small $e_1$};
\node (e2) at (2.5, -2.2) {\small $e_2$};
\end{scope}

\begin{scope}[every node/.style={steiner}]
\node (s1) at (-1.6, 3.2) {};
\node (s2) at (1.6, 3.2) {};
\node (s3) at (-1.6, 1.1) {};
\node (s4) at (1.6, 1.1) {};
\node (s5) at (-1.6, -1.1) {};
\node (s6) at (1.6, -1.1) {};
\node (s7) at (-1.6, -3.4) {};
\node (s8) at (1.6, -3.4) {};
\end{scope}

\begin{scope}[very thick]
\draw (a1) -- (s1);
\draw (a1) -- (s2);
\draw (s1) -- (b1);
\draw (s1) -- (b2);
\draw (s2) -- (c1);
\draw (s2) -- (c2);

\draw (s3) -- (b1);
\draw (s3) -- (b2);
\draw (s4) -- (c1);
\draw (s4) -- (c2);
\draw (a2) -- (s3);
\draw (a2) -- (s4);

\draw (a2) -- (s5);
\draw (a2) -- (s6);
\draw (s5) -- (d1);
\draw (s5) -- (d2);
\draw (s6) -- (e1);
\draw (s6) -- (e2);

\draw (s7) -- (d1);
\draw (s7) -- (d2);
\draw (s8) -- (e1);
\draw (s8) -- (e2);
\draw (a3) -- (s7);
\draw (a3) -- (s8);
\end{scope}

\begin{scope}[dual]
\begin{scope}[blue]
\draw (a1) ellipse (0.5 and 0.5);
\draw (0,2.55) ellipse (3.25 and 2.05);
\draw (a3) ellipse (0.5 and 0.5);
\draw (0,-3.7) ellipse (2.3 and 0.9);
\draw (0,-2.7) ellipse (3.35 and 2.15);
\end{scope}
\begin{scope}[red!80!black]
\draw (b1) ellipse (0.5 and 0.5);
\draw (b2) ellipse (0.5 and 0.5);
\end{scope}
\begin{scope}[green!60!black]
\draw (c1) ellipse (0.5 and 0.5);
\draw (c2) ellipse (0.5 and 0.5);
\end{scope}
\begin{scope}[orange]
\draw (d1) ellipse (0.5 and 0.5);
\draw (d2) ellipse (0.5 and 0.5);
\end{scope}
\begin{scope}[violet]
\draw (e1) ellipse (0.5 and 0.5);
\draw (e2) ellipse (0.5 and 0.5);
\end{scope}
\end{scope}

\end{scope}

\end{tikzpicture}
\end{center}
\caption{\label{fig:dual_solution_2}
The picture illustrates the values $y^r_{U,P}$ for $r=b_1$ (top), $r=a_1$ (middle) and $r=a_2$ (bottom) for an optimal dual solution for  $((V,E), c, \mathcal{P}_2)$.
The colored sets indicate non-zero variables $y^r_{U,P}$, where the color of the drawn set $U$ indicates the pair $P$.
For every drawn set $U$, we have a variable $y^r_{U,P}$ with value $1$, where \textcolor{blue}{$P=\{a_1,a_3\}$} for blue sets, 
\textcolor{red}{$P=\{b_1,b_2\}$} for red sets, \textcolor{green!60!black}{$P=\{c_1,c_2\}$} for green sets, \textcolor{orange}{$P=\{d_1,d_2\}$} for orange sets, and \textcolor{violet}{$P=\{e_1,e_2\}$} for violet sets.
}
\end{figure}

In this section we prove Theorem~\ref{thm:different_representaions} that we restate here for convenience.
\differentRepresentations*
\begin{proof}
The graph $G=(V,E)$ and the demand graphs for the two different sets $\Pscr_1$ and $\Pscr_2$ of pairs are shown in Figure~\ref{fig:different_representations}.
The cost $c(e)$ of every edge $e\in E$ is $1$.

In order to determine the value of \ref{eq:forest-bcr} for the two instances  $((V,E), c, \mathcal{P}_1)$ and $((V,E), c, \mathcal{P}_2)$, we will provide a pair of primal and dual LP solution for each of these instances. 
We remark that in order to prove Theorem~\ref{thm:different_representaions} it would be sufficient to provide a solution to \ref{eq:forest-bcr} for $((V,E), c, \mathcal{P}_1)$ and a solution to the dual of \ref{eq:forest-bcr} for $((V,E), c, \mathcal{P}_2)$ with strictly bigger value.
For convenience and to help intuition, we nevertheless provide both primal and dual solutions for both instances.

First, we provide primal solutions for the two different instances  $((V,E), c, \mathcal{P}_1)$ and $((V,E), c, \mathcal{P}_2)$ of value $12$ and $13$, respectively.
These solutions to \ref{eq:forest-bcr}  are shown in Figure~\ref{fig:different_primal_solutions}.

The dual of \ref{eq:forest-bcr} has variables $\alpha_P$ for each $P\in \mathcal{P}$ and variables $y^r_{U,P}$ for $r \in V$, $P\in \mathcal{P}$, and $ U \subseteq V \setminus \{ r \}$ with $P \cap U \neq \emptyset$.
It is given by:

\begin{align}\label{eq:dual-lp}\tag{Dual-Forest-BCR}
        \max &\ \sum_{P\in \mathcal{P}}  \alpha_P & \\
        \text{s.t.}\  \  &\ 
        \sum_{P\in \mathcal{P}} \ \sum_{\substack{U\subseteq V\setminus\{r\}: \\ P\cap U \neq \emptyset,\\ e \in \delta^+(U)}}  y^r_{U,P} \ \le\ c(e) & \text{ for all } r\in V \text{ and }e\in \overrightarrow{E}\nonumber \\[2mm]
       &\  \alpha_P \ \leq\ \sum_{\substack{U\subseteq V\setminus \{r\}: \\ P\cap U \neq \emptyset}} y^r_{U,P} &\text{ for all } P \in \mathcal{P}\text{ and }r\in V\nonumber \\[2mm]
       &\ y^r_{U,P} \ \geq\  0 &\text{ for all } r \in V , P\in \mathcal{P},\text{ and }\nonumber\\
        & &\  U \subseteq V \setminus \{ r \} \text{ with } P \cap U \neq \emptyset. \nonumber
\end{align}

We now construct a dual solution, i.e., a solution to \ref{eq:dual-lp}, for the instance $((V,E), c, \mathcal{P}_1)$ of value $12$.
To this end, we set $\alpha_P = 2$ for each pair $P\in \mathcal{P}_1$.
The values of the variables $y^r_{U,P}$ are illustrated in Figure~\ref{fig:dual_solution_1}.
The figure shows the values $y^r_{U,P}$ for $r \in \{ b_1, a_1, a_2\}$.
All other cases with $r\in R$ are symmetric. (The case $r\in\{b_2,c_1,c_2,d_1.d_2,e_1,e_2\}$ are symmetric to $r=b_1$and the case $r=a_3$ is symmetric to $r=a_1$.)
For the case where $r\in V\setminus R$ we can use the same values as for $r=b_1$  if $r$ is one of the Steiner vertices incident to $c_1$ and $c_2$, and  a symmetric version of this solution if $r$ is one of the other Steiner vertices.

Finally, we construct a dual solution, i.e., a solution to \ref{eq:dual-lp}, for the instance $((V,E), c, \mathcal{P}_2)$ of value $13$.
To this end, we set $\alpha_{\{a_1,a_3\}}=5$ and $\alpha_{\{a_1,a_2\}}=0$. 
Moreover, we set  $\alpha_P = 2$ for each pair $P\in \{\{b_1,b_2\},\{c_1,c_2\},\{d_1,d_2\},\{e_1,e_2\}\}$.
The values of the variables $y^r_{U,P}$ are illustrated in Figure~\ref{fig:dual_solution_2}.
The figure shows the values $y^r_{U,P}$ for $r \in \{ b_1, a_1, a_2\}$. 
Again all other cases with $r\in R$ are symmetric. 
For the case where $r\in V\setminus R$ we can use the same values as for $r=b_1$  if $r$ is one of the Steiner vertices incident to $c_1$ and $c_2$, and  a symmetric version of this solution if $r$ is one of the other Steiner vertices. 
This completes the proof.
\end{proof}

\newpage

\section*{Acknowledgment}
We thank the Schloss Dagstuhl
Leibniz-Zentrum für Informatik and the organizers of the Dagstuhl Seminar on "Parameterized Approximation: Algorithms and Hardness"(23291)'', during which the first idea of the rounding algorithm was developed.

\bibliographystyle{alpha}
\bibliography{BCR}

\end{document}